\newcommand*{\ATLASLATEXPATH}{latex/}
\documentclass[UKenglish,texlive=2011,cernpreprint]{\ATLASLATEXPATH atlasdoc}
\usepackage[subfigure=true]{\ATLASLATEXPATH atlaspackage}
\usepackage{\ATLASLATEXPATH atlasbiblatex}
\usepackage{\ATLASLATEXPATH atlascontribute}
\usepackage{\ATLASLATEXPATH atlasphysics}

\usepackage{enumitem}
\usepackage{amssymb,textcomp}

\graphicspath{{logos/}{figures/}}

\usepackage{paper-defs}

\AtlasTitle{Search for single production of vector-like quarks \\ decaying into $Wb$ in $pp$ collisions at $\sqrt{s} =$ 8 \TeV\ with the ATLAS detector}

\author{The ATLAS Collaboration}

\PreprintIdNumber{CERN-PH-EP-2015-319}

\AtlasJournal{\EPJC}

\AtlasAbstract{%
   A search for singly produced vector-like $Q$ quarks, 
   where $Q$ can be either a $T$ quark with charge $+2/3$ or a $Y$ quark with charge $-4/3$, 
   is performed in proton--proton collisions        
   recorded with the ATLAS detector at the LHC. The dataset corresponds
   to an integrated luminosity of 20.3 \ifb\ and was produced with a centre-of-mass energy of $\sqrt{s}=8$ \TeV.
   This analysis targets $Q \to Wb$ decays where the $W$ boson decays leptonically.
   A veto on massive large-radius jets is used to reject the dominant $t\bar{t}$ background. 
   The reconstructed $Q$-candidate mass, ranging from 0.4 to 1.2 \TeV, is used 
   in the search to discriminate signal from background processes. 
   No significant deviation from the Standard Model expectation is observed, and limits are set on the $Q \to Wb$ cross-section 
   times branching ratio. The results are also interpreted as limits on the $QWb$ coupling and the mixing with the Standard Model sector
   for a singlet $T$ quark or a $Y$ quark from a doublet.
   $T$ quarks with masses below 0.95 \TeV\ are excluded at 95\% confidence level,
   assuming a unit coupling and a BR$(T\rightarrow Wb) = 0.5$,
   whereas the expected limit is 1.10 \TeV.
}

\usepackage{xspace}

\def\herwig{\textsc{Herwig}\xspace}
\def\pythia{\textsc{Pythia}\xspace}

\def\alpgen{\textsc{Alpgen}\xspace}

\hypersetup{pdftitle={ATLAS document},pdfauthor={The ATLAS Collaboration}}

\begin{document}

\maketitle

\tableofcontents

\section{Introduction}
\label{sec:intro}
Despite the success of the Standard Model (SM) of particle physics at
energies up to the electroweak scale and its recent completion with the
discovery of a Higgs boson at the Large Hadron Collider~\cite{:2012gk,:2012gu}, 
it fails to describe phenomena such as the fermion mass hierarchy, 
the baryon asymmetry and the fine-tuning problem~\cite{Susskind:1978ms}. 
The existence of heavy vector-like quarks~\cite{delAguila:1982fs}
would allow for the cancellation of quadratic divergences that occur in loop corrections to the Higgs-boson mass,
solving the fine-tuning problem.
Vector-like quarks are defined as coloured (under SU(3)$_{\mathrm{c}}$ ) fermionic states that have left-handed and right-handed components 
that both transform in the same way in the SM gauge group
and therefore their masses are not obtained by a Yukawa coupling to the Higgs boson.
Their existence is, for example, predicted in Little Higgs models~\cite{ArkaniHamed:2002qy,Perelstein:2003wd,Perelstein:2005ka}, 
top-colour assisted technicolour~\cite{Hill:1994hp,Harris:1999ya,Harris:2011ez} or composite Higgs 
models ~\cite{Kaplan:1983fs,Kaplan:1983sm,Georgi:1984ef,Banks:1984gj,Georgi:1984af,Dugan:1984hq,Georgi:1985hf,Bellazzini:2014yua}. 

\begin{figure}[htbp]
 \centering
  \includegraphics[width=0.45\textwidth]{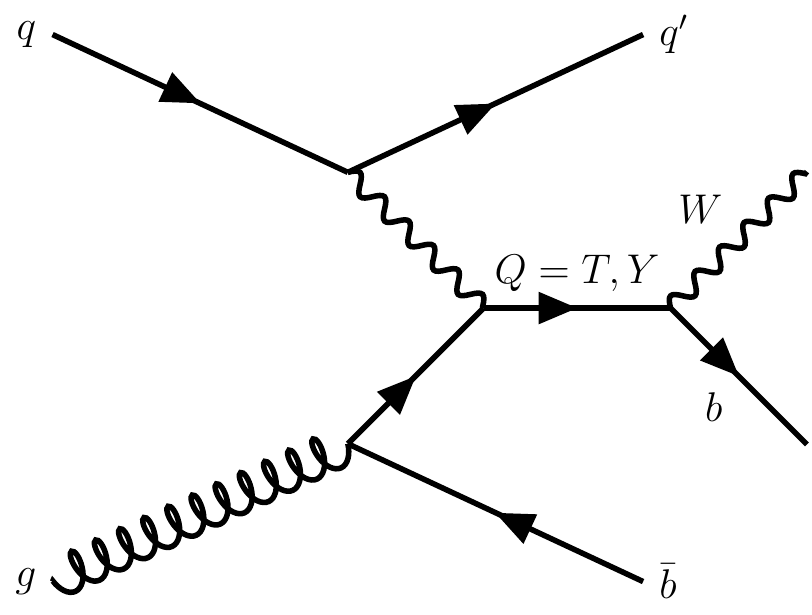}
  \caption{Leading-order Feynman diagram of single $Q=T,Y$ production and decay into $Wb$.}
  \label{fig:FeynmanDiagram}
\end{figure}

In this paper, a search for single production of heavy vector-like $Q$ quarks decaying into $Wb$ is presented. 
An example of a leading-order (LO) Feynman diagram is shown in Figure~\ref{fig:FeynmanDiagram}.  
The search targets the process $pp \rightarrow qQb$ with subsequent $Q \rightarrow Wb$ decay, 
where $Q$ can be either a $T$ quark with charge $+2/3$ or a $Y$ quark with charge $-4/3$.   
Heavy exotic fermions, such as vector-like quarks, are added to the SM in isospin multiplets.
\mbox{$T$ quarks} can belong to any multiplet, while $Y$ quarks cannot exist as singlets. 
The interpretation used in this paper focuses on $Y$ quarks from a $(Y,B)$ doublet and on singlet $T$ quarks.
For such $T$ quarks, the branching ratios (BRs) for $T$ are model dependent and mass dependent, 
but in the high-mass limit converge towards 2:1:1 ($Wb$:$Zt$:$Ht$).
The $Y \rightarrow Wb$ BR is 100\%.

The single production of vector-like quarks is enabled by their coupling to the SM quarks.
At higher masses, single production can become the dominant production process at the LHC 
depending on the strength of this coupling.
This dependence requires an interpretation of the results that relies on the formulation of the Lagrangian embedding these new interactions.
In this paper, two such interpretations are pursued, namely that in Ref.~\cite{Aguilar-Saavedra:2013qpa}
where a mixing term between the SM and vector-like quarks is introduced in a renormalisable extension of the SM, 
and in Refs.~\cite{DeSimone:2012fs,Matsedonskyi:2014mna} which uses a phenomenological Lagrangian parameterised
with coupling terms but which, however, is non-renormalisable.
When considering the phenomenology of these approaches,
the two main differences are the additional terms allowed in Refs.~\cite{DeSimone:2012fs,Matsedonskyi:2014mna},
which allow for larger production cross-sections,
and the complete description of the multiplet-dimension dependence of the BR in Ref.~\cite{Aguilar-Saavedra:2013qpa}.
The formulation of Ref.~\cite{Aguilar-Saavedra:2013qpa} also implies sensitivity to indirect electroweak constraints,
such as the ratio $R_{b}$ of the partial width for $Z \rightarrow b\bar{b}$ to the total hadronic $Z$-boson width
and the oblique parameters S and T~\cite{Peskin:1990zt}.

In this paper, the interpretation of the search for the single production of vector-like quarks  
is presented in terms of $\sin \theta$ and $c^{Wb}$, corresponding to the mixing and coupling terms introduced
by Ref.~\cite{Aguilar-Saavedra:2013qpa} and Refs.~\cite{DeSimone:2012fs,Matsedonskyi:2014mna}, respectively.
A comparison of their respective Lagrangians yields a simple relation\footnote{This relationship is only true within the regime of validity of the renormalisable formulation, and if one considers only the interactions between $Q$,$W$ and $b$.} between $\sin \theta$ and $c^{Wb}$ given by $c^{Wb}=\sqrt{2}\sin \theta$.
For the interpretation in terms of $c^{Wb}$, assumptions must be made about the $Q \rightarrow Wb, Q \rightarrow Zt$ 
and $Q \rightarrow Ht$ BRs, whereas $\sin\theta$ fully determines those BRs for any given heavy quark mass.
Therefore, in this paper, both interpretations are presented independently.
The relative contribution of the left- and right-handed components of the mixing and coupling
also depends on the dimension of the multiplet.
For $T$ singlets, the left-handed components (\sinL and \cL) are dominant.
For $Y$ quarks from a doublet, results are presented in terms of the magnitude of the total coupling $\sqrt{\cTwoL+\cTwoR}$,
while for the interpretation in terms of mixing, this can be simplified to just the contribution of 
the right-handed (\sinR) component~\cite{Aguilar-Saavedra:2013qpa}.

The ATLAS and CMS collaborations have published searches for pair-production of vector-like $T$ quarks
in all decay channels \cite{Aad:2015kqa,Aad:2014efa,Khachatryan:2015axa,Chatrchyan:2013uxa,Khachatryan:2015oba,Aad:2015gdg}. 
The best observed limits on the $T$-quark mass are $m(T)> 0.855$ \TeV\ for $Ht$~\cite{Aad:2015kqa}, 
0.810 \TeV\ for $Zt$~\cite{Aad:2014efa} and 0.920 \TeV\ for $Wb$~\cite{Khachatryan:2015oba} 
decay channels at the 95\% confidence level (CL), 
where a BR of 100\% is assumed to the corresponding decay channel.
For single $T$-quark production, searches for $T$ quarks with decays into $Zt$~\cite{Aad:2014efa} have been carried out 
by the ATLAS Collaboration using the 8 \TeV\ dataset, but for the $T \rightarrow Wb$ decay channel no mass limits have been set so far.

The analysis presented here is performed in the lepton+jets channel, 
characterised by the presence of exactly one electron or muon, and two or more jets.
The outgoing light quark in the process depicted in Figure~\ref{fig:FeynmanDiagram} 
typically produces a jet in the forward region of the detector.
One of the jets is a $b$-jet originating from the $Q$ decay. 
The $b$-jet and the charged lepton are back-to-back in the transverse plane since both originate from the decay of a heavy object.
The second $b$-jet originates from the gluon splitting and may be observed in either the forward or central region.
Since this $b$-jet is soft, it often falls outside the detector acceptance. 
The dominant backgrounds are $W$+jets, top-quark pair and single top-quark production. 
At higher \pT of top quarks and $W$ bosons, their decay products are more collimated.
They can be identified as one high-mass jet with a large radius parameter ($R$). 
Events with high-mass large-$R$ jets are vetoed to improve the suppression of the large \ttbar$\rightarrow WbWb$ background process
where one $W$-boson decays hadronically and the other leptonically.

\section{ATLAS detector}

\label{Sec:ATLASDetector}

The ATLAS detector~\cite{atlas-detector} is a forward-backward symmetric multi-purpose detector 
and covers almost the full solid angle.\footnote{ATLAS uses a right-handed coordinate system with its origin at the nominal 
interaction point (IP) in the centre of the detector and the $z$-axis along the beam pipe. 
The $x$-axis points from the IP to the centre of the LHC ring, and the $y$-axis points upward. 
Cylindrical coordinates $(r,\phi)$ are used in the transverse plane, $\phi$ being the azimuthal angle around the beam pipe. 
The pseudorapidity is defined in terms of the polar angle $\theta$ as $\eta=-\ln\tan(\theta/2)$,
and the distance between two objects in $\eta$-$\phi$ space is measured in terms of 
$\Delta R \equiv \sqrt{(\Delta\eta)^{2}+(\Delta\phi)^{2}}$} 
The inner detector (ID) is installed closest to the beam pipe, covering the pseudorapidity range $|\eta| <$ 2.5. 
The ID comprises a silicon pixel detector and a silicon microstrip detector up to \mbox{$|\eta| <$ 2.5} 
and a transition radiation tracker up to $|\eta| <$ 2.0. 
The ID is immersed in an axial 2 T magnetic field provided by a superconducting solenoid.
Outside the solenoid magnet is the electromagnetic liquid-argon (LAr) sampling calorimeter, 
which has high granularity and covers up to $|\eta| =$ 3.2. 
The central part of the hadronic calorimeter (up to $|\eta| <$ 1.7) uses scintillator tiles as the active medium, 
while the forward part is a sampling calorimeter using LAr \mbox{(1.5 $< |\eta| <$ 4.9).} 
The outer part of the ATLAS detector is the three-layer muon spectrometer which is immersed in a magnetic field
provided by a large air-core toroid system. 

The muon tracks are measured in $|\eta| <$ 2.7 using monitored drift tubes and cathode-strip chambers, 
while resistive-plate and thin-gap chambers are used in the trigger system for $|\eta| <$ 2.4.

Events are selected using a three-level trigger system~\cite{Aad:2012xs}. In the first step (Level-1), 
the event rate is reduced to 75 kHz using hardware-based triggers. 
The High-Level Trigger (Level-2 and Event Filter) is software based and reduces the rate to 400 Hz. 

\section{Data and simulation samples}
\label{Sec:Samples}
The search presented in this paper uses $pp$ collision data at $\sqrt{s} =$ 8 \TeV\ that were collected with the ATLAS detector in 2012. 
The data used for this analysis were taken under stable beam conditions 
and with all relevant ATLAS subdetector systems operational.
The integrated luminosity of the data sample corresponds to 20.3 $\pm$ 0.6 fb$^{-1}$~\cite{Aad:2013ucp}. 
The events were selected using single-electron and single-muon triggers. 
Monte Carlo (MC) samples are generated in order to model the signal and background processes. 
In the MC simulation, multiple $pp$ interactions in the same and neighbouring bunch crossings (pile-up) are taken into account.
A weighting procedure is used to correct the simulated events such that they have the same pile-up distribution as the data.
{\scshape Geant4}~\cite{Geant4} is used to simulate the full ATLAS detector~\cite{Sim} for the generated data. 
The simulated events and the ATLAS data are processed with the same reconstruction software.

The signal MC samples are based on the model described in Ref.~\cite{Vignaroli:2012nf} 
and are generated with {\scshape MadGraph} v5 \cite{Alwall:2011uj} using a UFO model~\cite{Degrande:2011ua,Alloul:2013bka} 
and the CTEQ6L1 parton distribution functions (PDFs)~\cite{cteq6}. 
The samples are generated in the $t$-channel using the \mbox{$2 \to 3$} process $pp \to qQb$,
with $Q$ decaying exclusively into $Wb$ and $W$ decaying inclusively into all the available modes. In the case that a branching ratio of 50\% is used, the corresponding signal yields are scaled by a factor of 0.5. Other decay modes of $Q$ are assumed to be negligible and are not taken into account.
The events are interfaced with {\scshape Pythia8}~\cite{Sjostrand:2007gs} for parton showering, hadronisation and particle decay. 
Signal samples are generated with different $Q$ masses in the range 0.4--1.2 \TeV\ in steps of 0.1 \TeV. 
All signal samples are produced using the narrow-width approximation with a width of $\Gamma/m = 7\%$. 
Additional samples with $\Gamma/m$ varying from 2\% to 46\% are used to examine the dependence of the vector-like quark width on \cL. 

The dominant backgrounds are \ttbar, $W$+jets and single top-quark production. Smaller background contributions are $Z$+jets, 
diboson and multijet production. 
The \ttbar and single top-quark processes are modelled using the next-to-leading-order (NLO) {\scshape Powheg-Box}
generator r2330.3~\cite{Frixione:2007vw} using the CT10 PDFs~\cite{Lai:2010vv}. 
{\scshape Powheg-Box} is then interfaced with {\scshape Pythia} v6.4~\cite{Sjostrand:2006za} with the Perugia 2011C set of tuned parameters~\cite{Skands:2010ak} 
and the CTEQ6L1 PDFs. The top-quark mass is set to 172.5 \GeV\ in all samples. 
The \alpgen v2.13~\cite{Mangano:2002ea} LO generator and the CTEQ6L1 PDF set are used to simulate $W$/$Z$ production.
Parton showers and hadronisation are modelled with \pythia v6.4.   
The $W$/$Z$ samples are generated with up to five additional partons, separately for $W$/$Z$+light-jet,
$W$/$Z+b\bar{b}$, $W$/$Z+c\bar{c}$ and $Wc$.
To avoid double-counting of partonic configurations generated by both the matrix-element calculation and
the parton-shower evolution, a parton–jet matching scheme (MLM matching)~\cite{Mangano:2001xp} is employed.   
The overlap between $W$/$Z+q\bar{q}$ ($q=b,c$) events generated from the matrix-element calculation and those
generated from parton-shower evolution in the $W$/$Z$+light-jet samples is avoided via an algorithm based
on the distance in $\eta-\phi$ space between the heavy quarks: if $\Delta R (q,\bar{q}) > 0.4$,
the matrix-element prediction is used, otherwise the parton-shower prediction is used.
Diboson samples with at least one leptonically-decaying boson are produced using \herwig v6.52~\cite{Corcella:2000bw} 
and {\scshape Jimmy} v4.31~\cite{Butterworth:1996zw} using the CTEQ6L1 PDFs. Multijet production is modelled from data as described later.

A control region is used to obtain the normalisations and corresponding uncertainties for the \ttbar and $W$+jets contributions. 
Theoretical calculations of cross-sections are used to normalise the predictions of the smaller backgrounds. 
The inclusive $Z$+jets cross-section is calculated to next-to-next-to-leading-order (NNLO) accuracy using FEWZ~\cite{Anastasiou:2003ds}. 
The single top-quark production cross-sections are calculated at NLO+NNLL (next-to-next-to-leading-logarithmic) precision in QCD. 
The largest contribution comes from $t$-channel production, 
with a corresponding uncertainty of +3.9/$-$2.2\% \cite{Kidonakis:2011wy}.
Additional samples are generated to model the systematic uncertainties of the dominant backgrounds. 
The effect of initial-state radiation (ISR) and final-state radiation (FSR) on the \ttbar\ background 
is estimated using the LO {\scshape AcerMC} v3.8~\cite{Kersevan:2004yg} generator interfaced with \pythia v6.4 and using the CTEQ6L1 PDFs. 
A measurement of $t\bar{t}$ production with a veto on additional central jet activity~\cite{ATLAS:2012al} 
is used to determine the ranges within which the parameters related to ISR and FSR are varied in \pythia.

The effect of using different models for hadronisation and factorisation is taken into account with
a sample generated with {\scshape Powheg-Box} but interfaced to {\scshape Herwig} v6.52 
using the CT10 PDFs in the matrix-element. 
The uncertainty due to the choice of \ttbar generator is modelled 
by comparing the default sample to a {\scshape MC@NLO} v4.03~\cite{Frixione:2002ik,Frixione:2003ei} sample interfaced with {\scshape Herwig} v6.52
using the CT10 PDF set 
and a sample produced with the multi-parton generator {\scshape Alpgen}+{\scshape Herwig} v6.52 (with up to three additional jets) 
using the CTEQ6L1 PDFs. 
For the evaluation of the single-top-quark modelling uncertainty, the default $t$-channel sample is compared to a sample generated with {\scshape MadGraph5\_}a{\scshape MC@NLO}~\cite{Alwall:2014hca} and {\scshape Herwig} v6.52 using the CT10 PDF set.

\section{Object definition}
\label{Sec:ObjectDefinition}

The search for vector-like $Q$ quarks and the reconstruction of the $Q$-candidate mass relies on the identification of jets, 
electrons, muons and missing transverse momentum \MET. 
Jets are reconstructed with the anti-$k_{t}$ algorithm~\cite{Cacciari:2008gp} 
with radius parameters of $R =$ 0.4 (small-$R$ jets) and $R =$ 1.0 (large-$R$ jets). 
Locally calibrated topological clusters of calorimeter cells~\cite{Lampl:1099735,Aad:2011he} 
are calibrated to the energy scale of particle-level hadrons and are used as input to the jet clustering algorithm. 
Small-$R$ jets are required to have a \pT\ greater than 25 \GeV\ for $|\eta| <$ 2.4,
while for forward jets, with 2.4 $< |\eta| <$ 4.5, \pT $>$ 35 \GeV\ is required.
The higher jet \pT\ threshold for forward jets is used to mitigate pile-up effects.
Large-$R$ jets are required to have \pT $>$ 200 \GeV\ and $|\eta| <$ 2.0. 
To reduce the influence of pile-up and of soft QCD radiation on large-$R$ jets a trimming procedure is used \cite{Krohn:2009th}, 
where the jet constituents are clustered into subjets using the $k_{t}$ algorithm~\cite{Catani:1993hr} with $R =$ 0.3. 
These subjets are removed from the large-$R$ jet if they fulfil $p_\mathrm{T}^{\rm subjet} < 0.05 \, p_\mathrm{T}^{\operatorname{\mathrm{large-}{\it R}}\, \mathrm{jet}}$
and the kinematics of the large-$R$ jet are recalculated.

In order to further suppress jets originating from pile-up, 
a requirement on the jet vertex fraction (JVF)~\cite{Aad:2015ina} is made. 
The JVF is defined as the summed scalar \pT of tracks associated 
with both the reconstructed primary vertex and the small-$R$ jet, 
divided by the summed scalar \pT of all tracks associated with the jet.
For jets with \pT $<$ 50 \GeV\ and |\eta| $<$ 2.4, a JVF $\geq$ 0.5 is required. 
When the small-$R$ jets are built, the jets and electrons are not distinguished. 
Hence, an electron will also be reconstructed as a jet. 
To remove these objects, the jet closest to a selected electron is removed if $\Delta R({\mathrm{jet}},e) <$ 0.2.

Jets containing $b$-hadrons are identified ({\it $b$-tagged})
using properties specific to these hadrons, such as a long lifetime and a large mass.
This analysis uses a multivariate discriminant~\cite{Aad:2015ydr} that is based
on displaced vertices and the impact parameters
of tracks associated with the jet.
The algorithm has an efficiency of 70\% to select $b$-jets and rejection factors of 5 and 135 for $c$-jets and light-quark or gluon jets, respectively, when assessed in a \ttbar simulated sample.

To reconstruct electrons, ID tracks are matched to energy deposits in the electromagnetic calorimeter~\cite{Aad:2014nim,Aad:2014fxa}. 
Only electrons with a transverse energy, $E_\mathrm{T} = E_{\rm cluster} / \cosh (\eta_{\rm track})$, greater than 25 \GeV\ are considered 
in the analysis. The \pT\ threshold of the offline lepton is higher than the momentum threshold of the trigger 
to ensure a trigger efficiency that is uniform in \pT\ for the selected leptons.
The energy cluster of the electron candidate must have a pseudorapidity $|\eta_{\rm cluster}| <$ 2.47. 
Electrons in the transition region between the barrel calorimeter and the endcap calorimeter  (1.37 $\leq |\eta| \leq$ 1.52) are rejected. 
To reject electrons originating from heavy-flavour hadron decays, 
electrons within a cone of size $\Delta R = $ 0.4 around a jet are removed from the event. 
For calorimeter-based isolation, a requirement on the energy deposited 
in clusters within a $\Delta R = $ 0.2 cone around the electron is made.
The energy of the electron is subtracted and pile-up corrections are applied. 
A similar procedure is used for track-based isolation, using $\Delta R=0.3$.
Calorimeter-based and track-based isolation criteria which are dependent on $E_{\mathrm T}$ and $\eta$ 
ensure 90\% isolation efficiency at all electron \pT values
for \mbox{$\Delta R =$ 0.2} and 0.3, respectively. 
A requirement on the longitudinal impact parameter $z_0$ is made to the electron track, requiring $|z_0| <$ 2 mm.

\clearpage

\subfigcapmargin = .5cm

\begin{figure}[!htb]
\centering
\subfigure[Leading large-$R$ jet \pT\ after preselection requirements described in Section~\ref{Sec:EventSelection}.]{\includegraphics[width=0.48\linewidth]{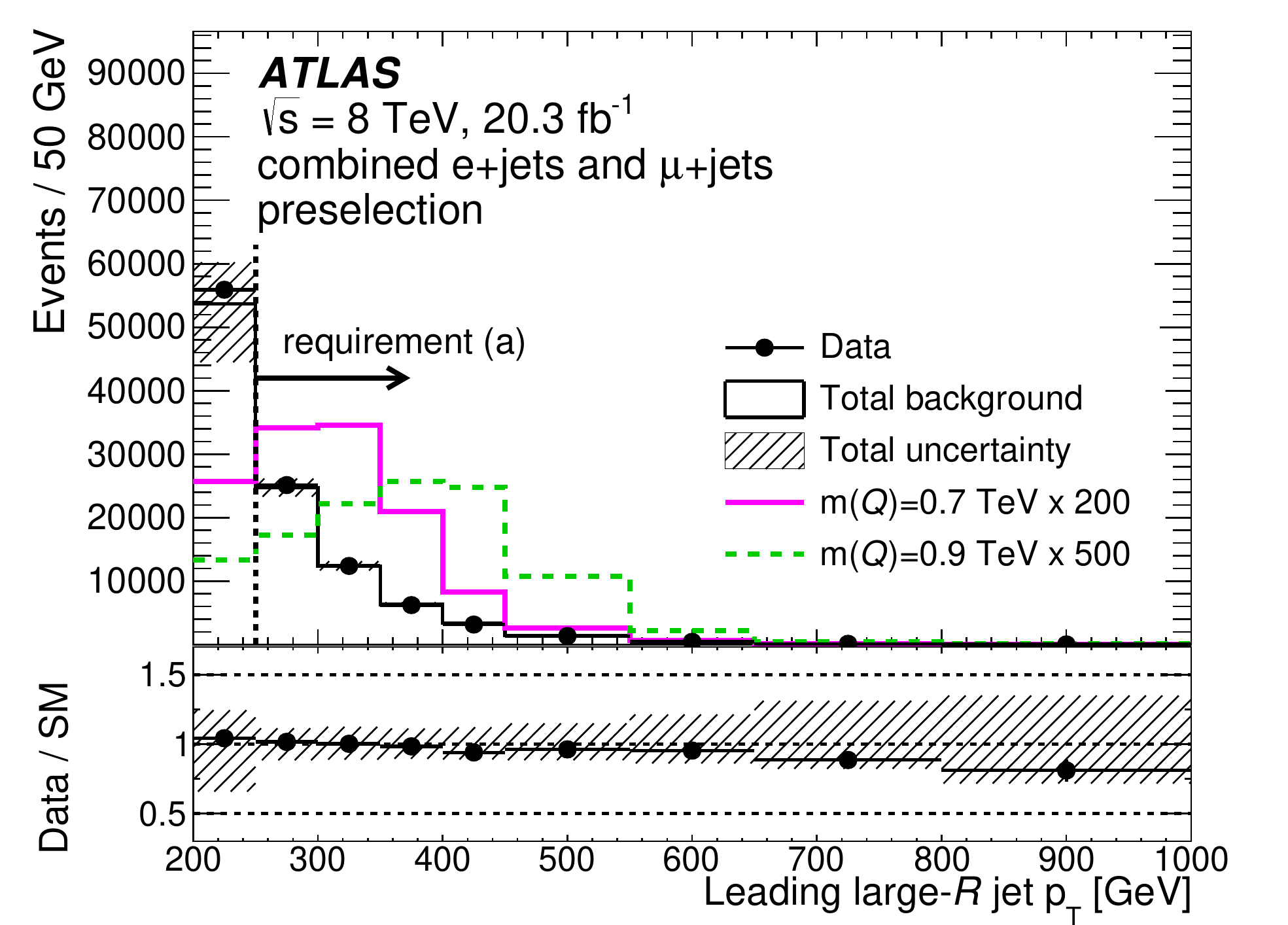}}
\subfigure[Leading large-$R$ jet mass after preselection and the requirement (a) described in Section~\ref{Sec:EventSelection}.]{\includegraphics[width=0.48\linewidth]{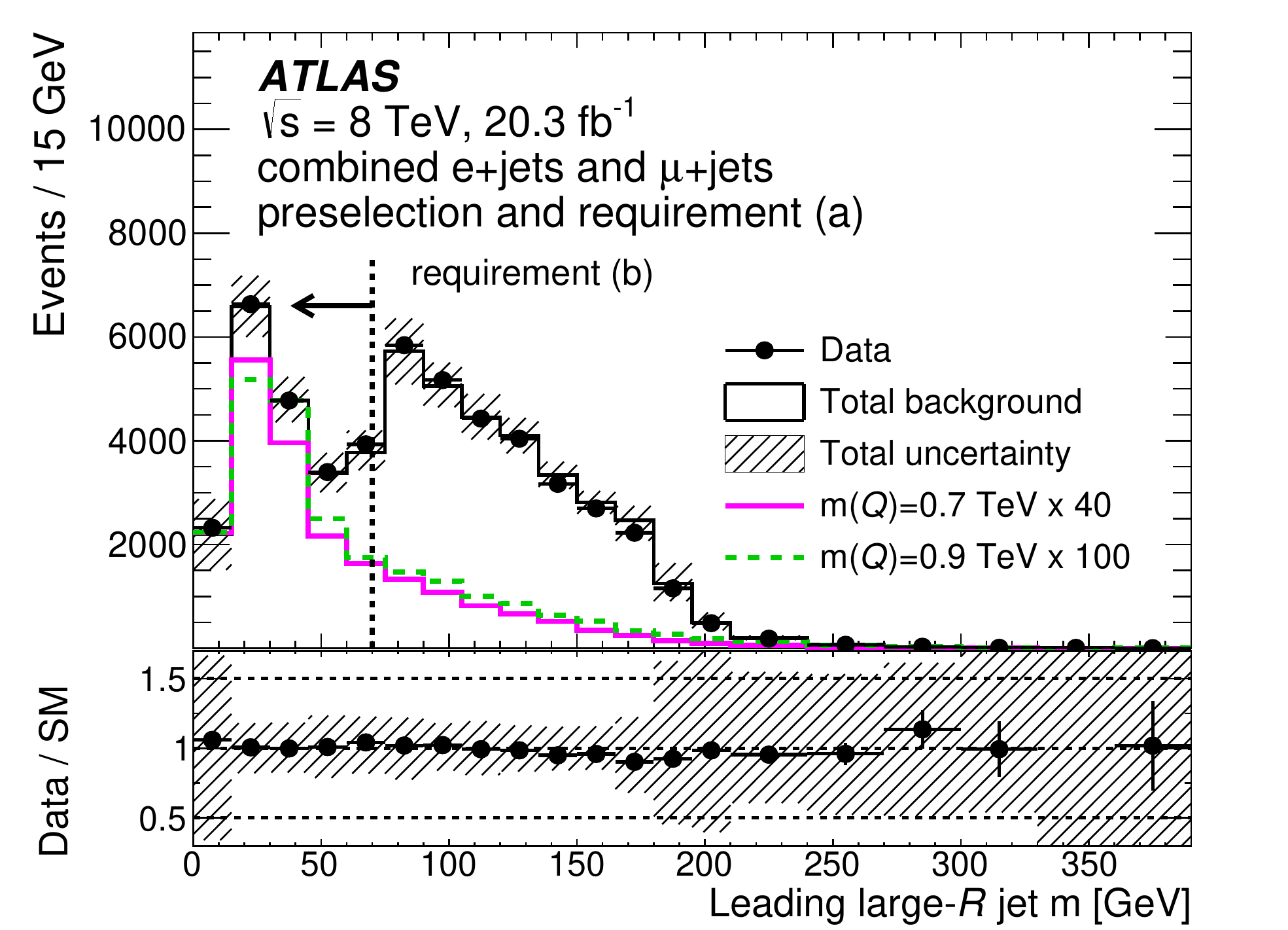}}
\subfigure[Number of $b$-tagged jets geometrically matched to the leading large-$R$ jet after preselection and requirements (a) and (b) described in Section~\ref{Sec:EventSelection}.]{\includegraphics[width=0.48\linewidth]{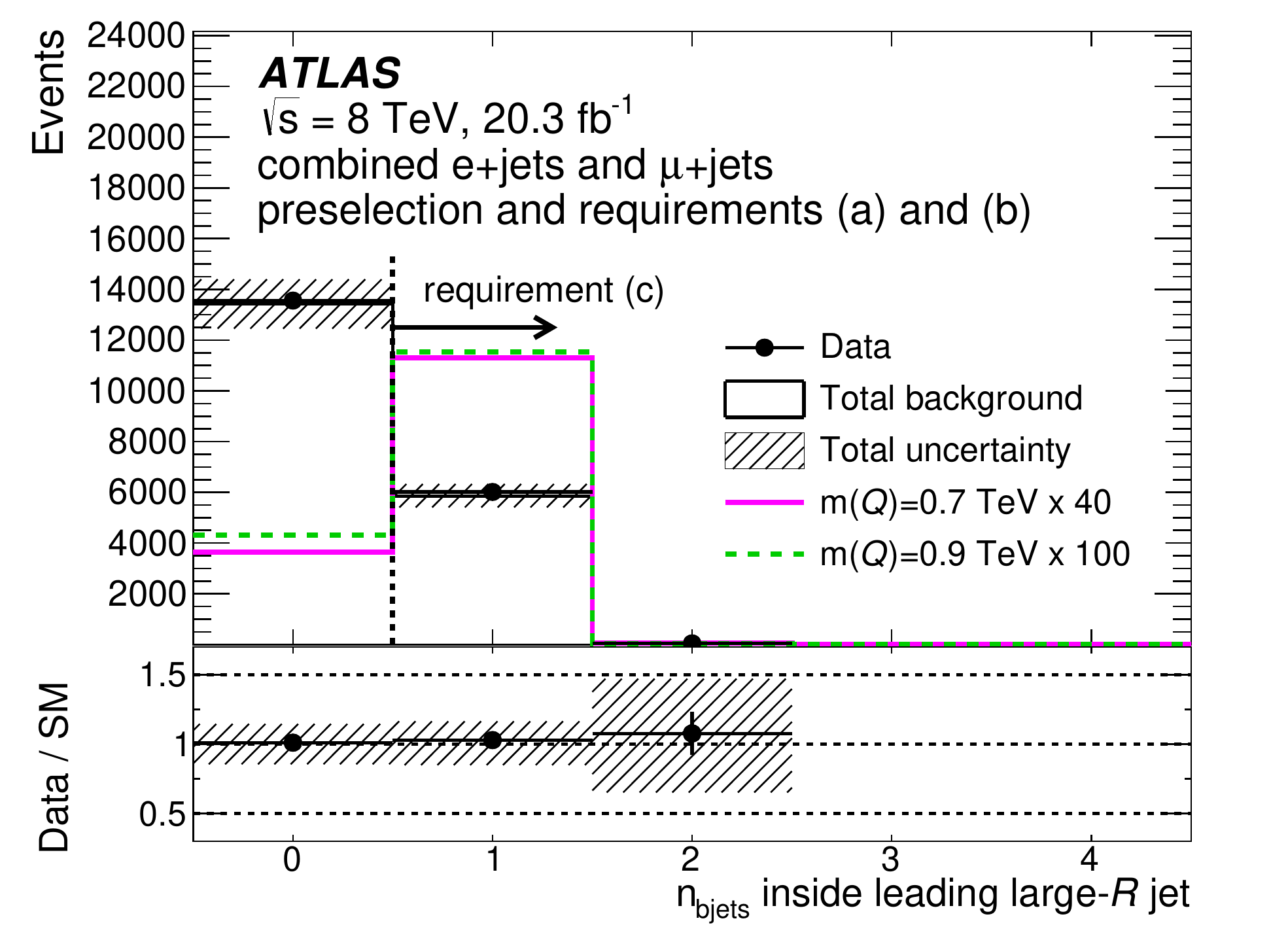}}
\subfigure[$\Delta \phi$ between the lepton and the large-$R$ jet after preselection and requirements (a)--(c) described in Section~\ref{Sec:EventSelection}.]{\includegraphics[width=0.48\linewidth]{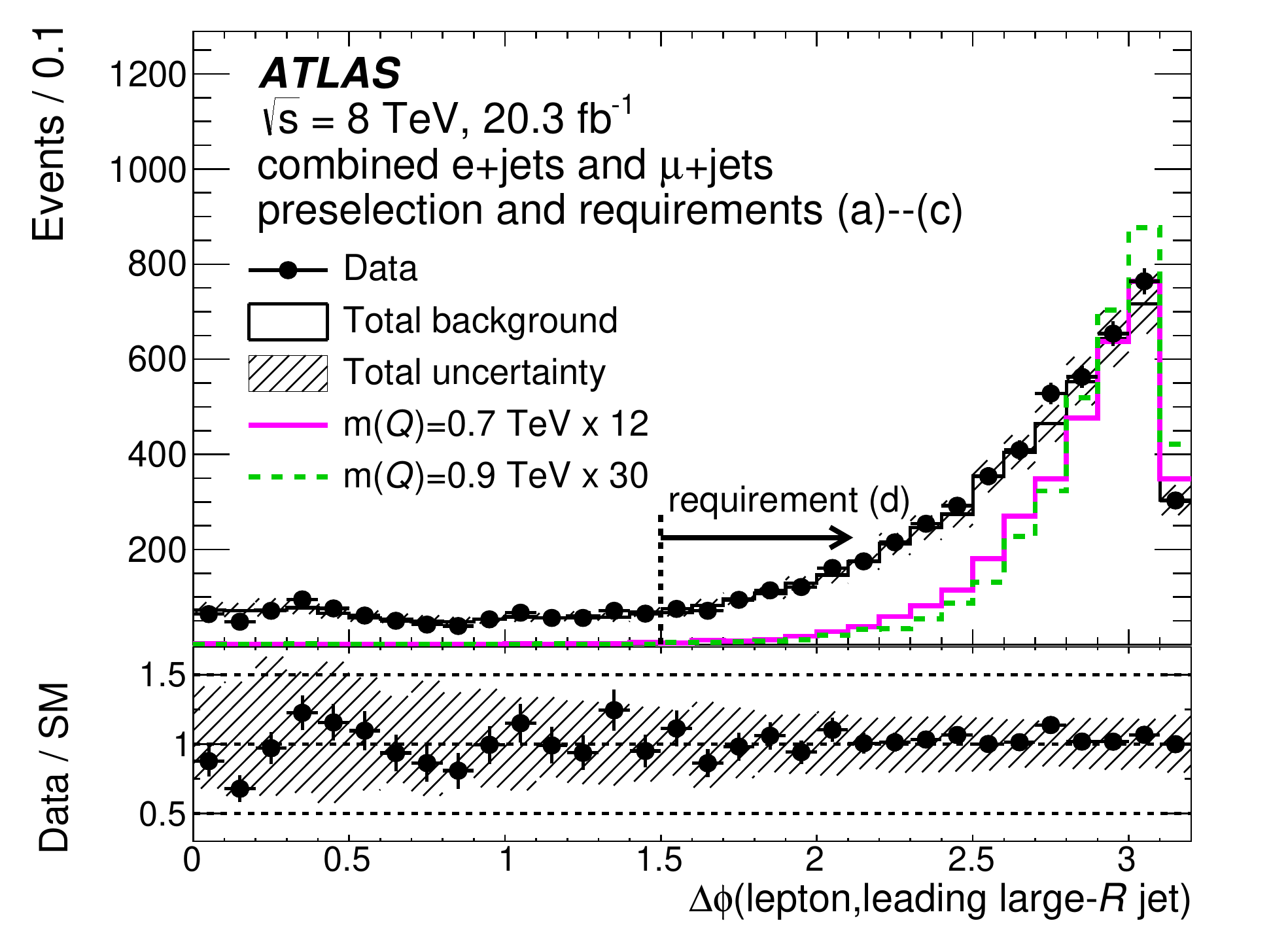}}
\subfigure[Number of jets outside the large-$R$ jet with $\pT \geq 75$ \GeV\ and $|\eta| \leq 2.4$ after preselection and requirements (a)--(d) described in Section~\ref{Sec:EventSelection}.]{\includegraphics[width=0.48\linewidth]{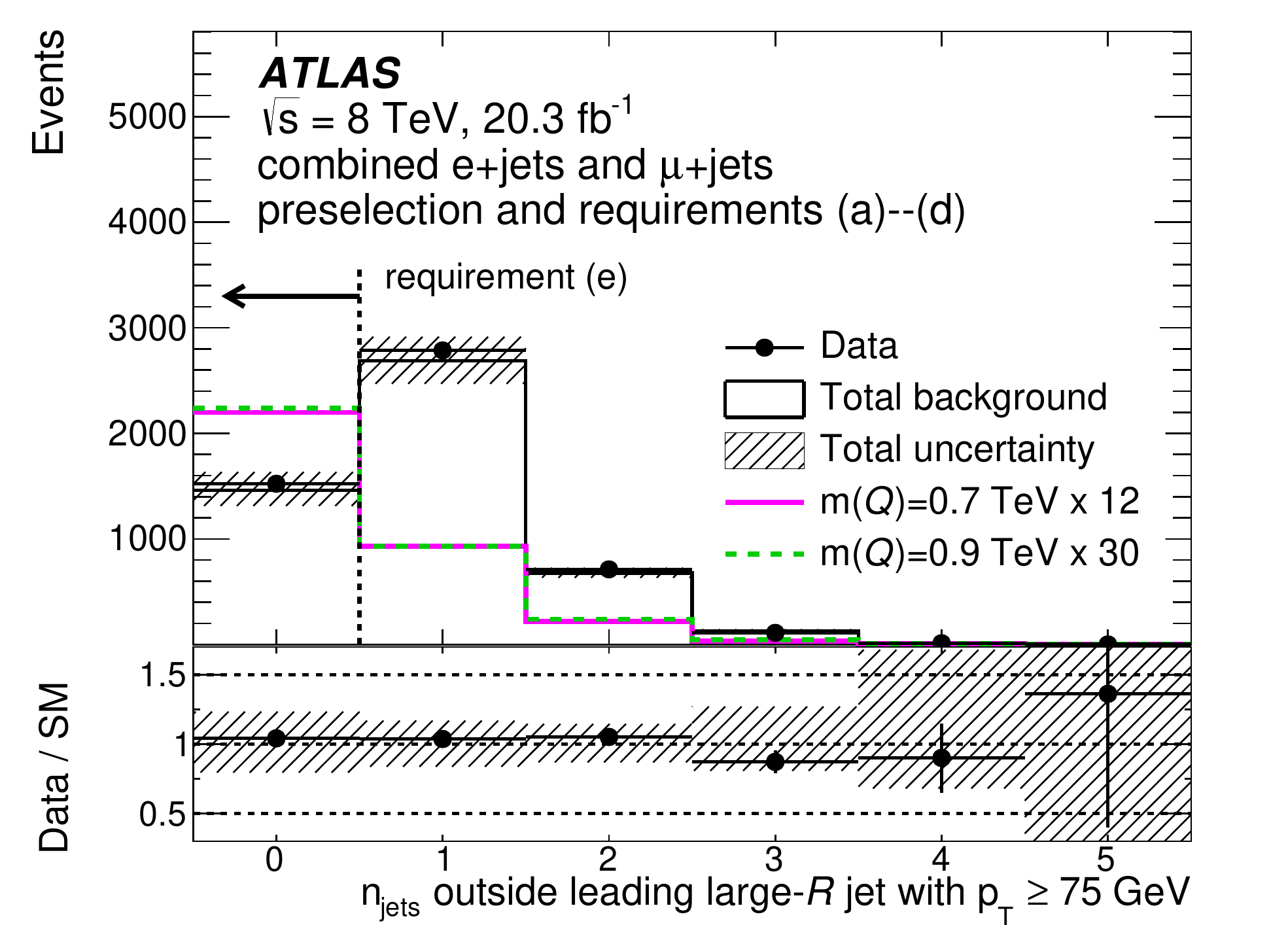}}
\subfigure[Number of forward jets after preselection and requirements (a)--(e) described in Section~\ref{Sec:EventSelection}.]{\includegraphics[width=0.48\linewidth]{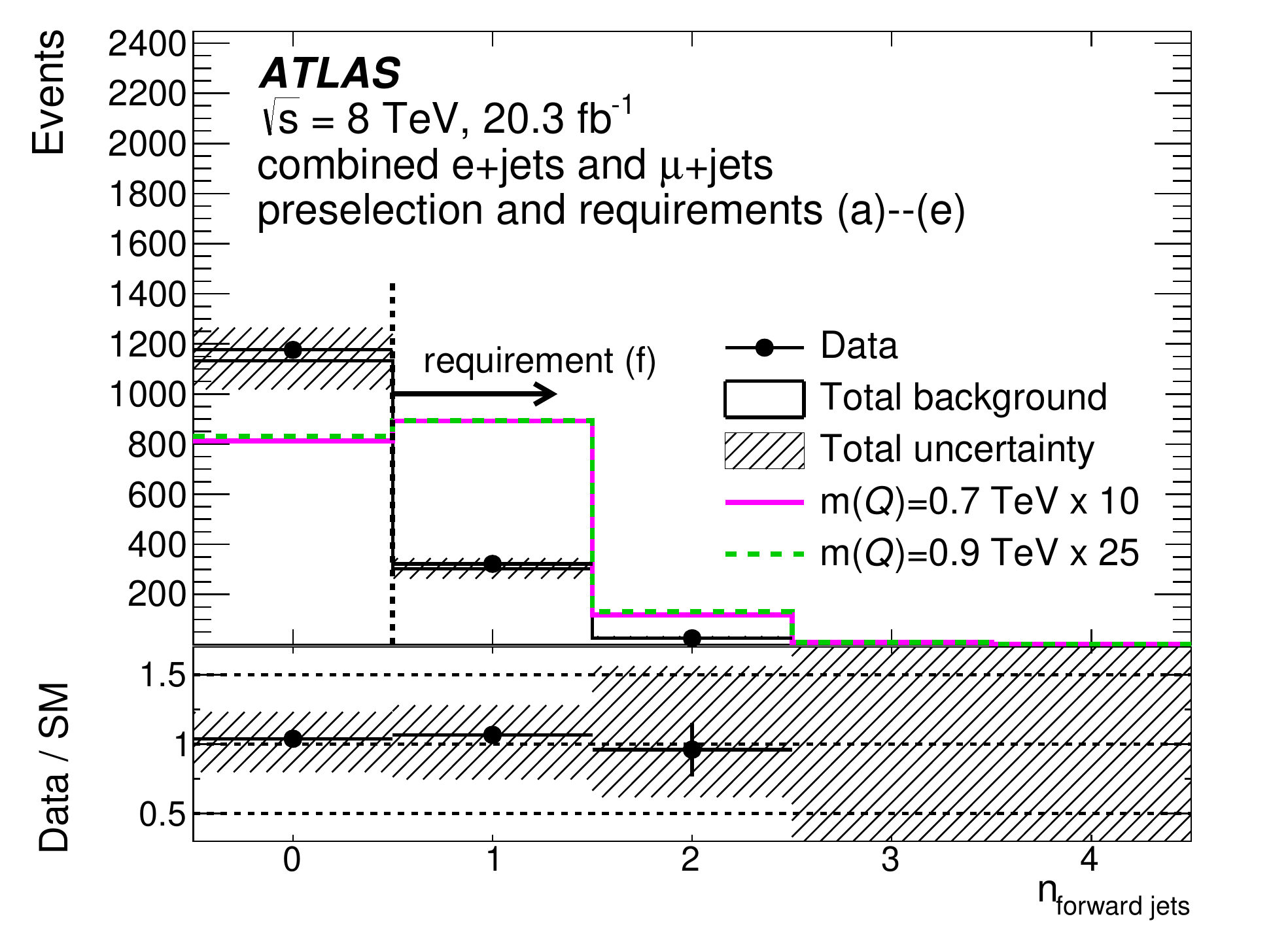}}
\caption{Comparison of data to expected background for the variables used in the event selection.
Each distribution is shown for events satisfying the preceding steps. 
The signal yields are shown for \cL = 1 and for BR($T \to  Wb$) = 0.5.
These are scaled up, in order to improve their visibility.
Scale factors are chosen to ease a shape comparison between the signal samples shown.
The distributions are shown here for the combined $e$+jets and $\mu$+jets channels.
}
\label{fig:CutFlow_6}
\end{figure}

\clearpage

For the identification of muons, tracks from the ID and the muon spectrometer are combined~\cite{Aad:2014rra}. 
Muons are required to have a \pT\ larger than 25 \GeV\ and $|\eta| <$ 2.5. 
Muons are required to be isolated from other high-\pT tracks within a small cone around the muon track. 
The size of the cone varies as a function of the muon \pT according to $\Delta R =$ 10 \GeV$/$\pT\cite{Rehermann:2010vq}.  
The muon is considered to be isolated if the scalar sum of the \pT from all other tracks in the cone is less than 5\% of the muon \pT.  
This requirement has an average efficiency of 97\%.
To reject muons originating from heavy-flavour decays, 
muons within a $\Delta R =$ 0.4 cone around a jet are removed. 
The longitudinal impact parameter of the muon track has to fulfil $|z_0| <$ 2 mm.

The neutrino from the leptonic $W$-boson decay cannot be observed directly, but its presence leads to \MET. 
To reconstruct the \vecMET, the vectorial sum of the momenta of all reconstructed electrons, 
muons and jets as well as all additional energy deposits in the calorimeters is calculated~\cite{Aad:2012re,ATLAS-CONF-2012-101}. 
The energy of clusters in the calorimeters matched to electrons, 
muons or jets is corrected according to the nature of the associated object.

\section{Event selection}
\label{Sec:EventSelection}

This section defines the signal region (SR) and control regions (CRs). 
The event selection presented here is based on the strategy proposed in Ref.~\cite{Ortiz:2014iza}. 
The preselection of events in the SR requires each event to have exactly one isolated lepton (electron or muon) 
as defined in Section~\ref{Sec:ObjectDefinition}. 
Furthermore, this lepton must be matched to the lepton that was reconstructed by the trigger. 
At least two small-$R$ jets and at least one large-$R$ jet are 
required; however, the large-R jet may contain one of the small-R jets.
\footnote{The small-$R$ jets and large-$R$ jets are clustered independently, using all available clusters in the calorimeter, 
therefore these objects can overlap.} 
The event must have a reconstructed primary vertex with at least five tracks with $p_{\mathrm T} >$ 400 \MeV. 
To suppress multijet background, the \MET needs to be larger than 20 \GeV\ and 
the sum of the \MET and the $W$-boson transverse mass,
$m_{\mathrm{T}} (W) = \sqrt{2 p_\mathrm{T}^{\ell} E_{\mathrm{T}}^{\mathrm{miss}} (1 - \cos \phi(\ell, \vec{E}_\mathrm{T}^{\mathrm{miss}}))}$,
must be larger than 60 \GeV. The angle between the transverse momentum of the lepton and the \vecMET vector 
is defined as $\phi(\ell, \vec{E}_\mathrm{T}^{\mathrm{miss}})$.

Several discriminating variables are used to further optimise the selection and define the SR.
These requirements are explained in the following.
Since $T$ quarks are excluded for masses below 0.7 \TeV, 
the optimisation of the selection criteria is done for the 0.7 \TeV\ mass point. 
The sequence of the final selection is illustrated in Figures~\ref{fig:CutFlow_6}(b)--~\ref{fig:CutFlow_6}(f), 
for the combined $e$+jets and $\mu$+jets channels,
following the order in which each criterion is applied.
After the preselection, the final sequence of requirements is:
\begin{enumerate}[nolistsep]
\item[(a)] The highest-\pT\ (leading) large-$R$ jet \pT\ must be greater than 250 \GeV.
\item[(b)] Events with massive large-$R$ jets ($m > 70$ \GeV) are rejected.
\item[(c)] At least one $b$-tagged jet matched to the large-$R$ jet, $\Delta R ($large-$R$ jet, $b$-tagged jet$)<0.8$, is required.
\item[(d)] The azimuthal separation between the lepton and the large-$R$ jet is required to be larger than 1.5.
\item[(e)] Events with any jet with $\pT > 75$ \GeV\ and $|\eta| < 2.4$ outside the large-$R$ jet are rejected.
\item[(f)] At least one forward jet is required in the event.
\end{enumerate}
For the MC signal samples used, the combined acceptance times efficiency
is 1.4\%\ for both \mbox{$m(T)=0.7$ \TeV} and $m(T)=0.9$ \TeV.

\section{Background estimation}

\label{Sec:Background}

The multijet background is obtained from data using a matrix method~\cite{ATLAS-CONF-2014-058}
which predicts the shape and normalisation of the background process. 
This method relies on differences between the probability of 
a ``real'' (prompt) lepton and that of a ``fake'' (non-prompt or misidentified) 
lepton to fulfil certain selection criteria. 
The ``fake'' lepton efficiencies are measured in data using background-enriched control regions 
and are parameterised for different values of \pT\ and $\eta$ of the charged lepton candidate. 
The ``real'' lepton efficiencies are measured in $Z \to \ell \ell$ samples containing prompt leptons.

All other background shapes are obtained from simulation, using the samples discussed in Section~\ref{Sec:Samples}. 
A fit control region (FitCR) is defined in order to estimate the normalisation of the \ttbar\ background and of 
the $W$+jets background from data. 
Two additional $W$+jets-enriched CRs are defined to validate the modelling (W1CR and W2CR). 

In order to suppress the $t\bar{t}$ contribution in the W2CR, the following requirement is made:
\begin{enumerate}[nolistsep,noitemsep]
\item[(g)] The invariant mass of the charged lepton and the $b$-tagged jet should be be larger than 175 \GeV.
\end{enumerate}
This requirement is not applied in any other region.
All CRs are orthogonal to the SR, which is achieved by inverting requirement (e) as defined in Section~\ref{Sec:EventSelection}. 
Therefore, instead of applying the jet veto, events are required to have a jet in that regime.
The relation between the requirements used to define these CRs and the SR are summarised in Table~\ref{table:table_appx_CRs}.

\begin{table}[htbp]
\centering
\caption{Differences in the event selections applied in the SR and CRs. A checkmark (\checkmark) is shown if the specific requirement 
is applied in the region, the cross ($\times$) shows that a requirement is not applied. 
Requirements (a) and (d) are applied in the SR and all CRs.}
\vspace{0.2cm}
\begin{tabular}[c]{ccccc}
\hline
\hline Requirements & SR & FitCR & W1CR & W2CR \\
\hline
(a) & \checkmark & \checkmark & \checkmark & \checkmark \\
(b) & \checkmark & {\bf $\times$} & \checkmark & \checkmark \\
(c) & \checkmark & \checkmark & \checkmark & {\bf $\times$} \\
(d) & \checkmark & \checkmark & \checkmark & \checkmark \\
(e) & \checkmark & inverted & inverted & inverted \\
(f) & \checkmark & {\bf $\times$} & {\bf $\times$} & {\bf $\times$} \\
(g) & {\bf $\times$} & {\bf $\times$} & {\bf $\times$} & \checkmark \\
\hline
\hline
\end{tabular}
\label{table:table_appx_CRs}
\end{table}

\begin{figure}[!htb]
  \centering
  \subfigure[$e$+jets channel]{\includegraphics[width=0.49\textwidth]{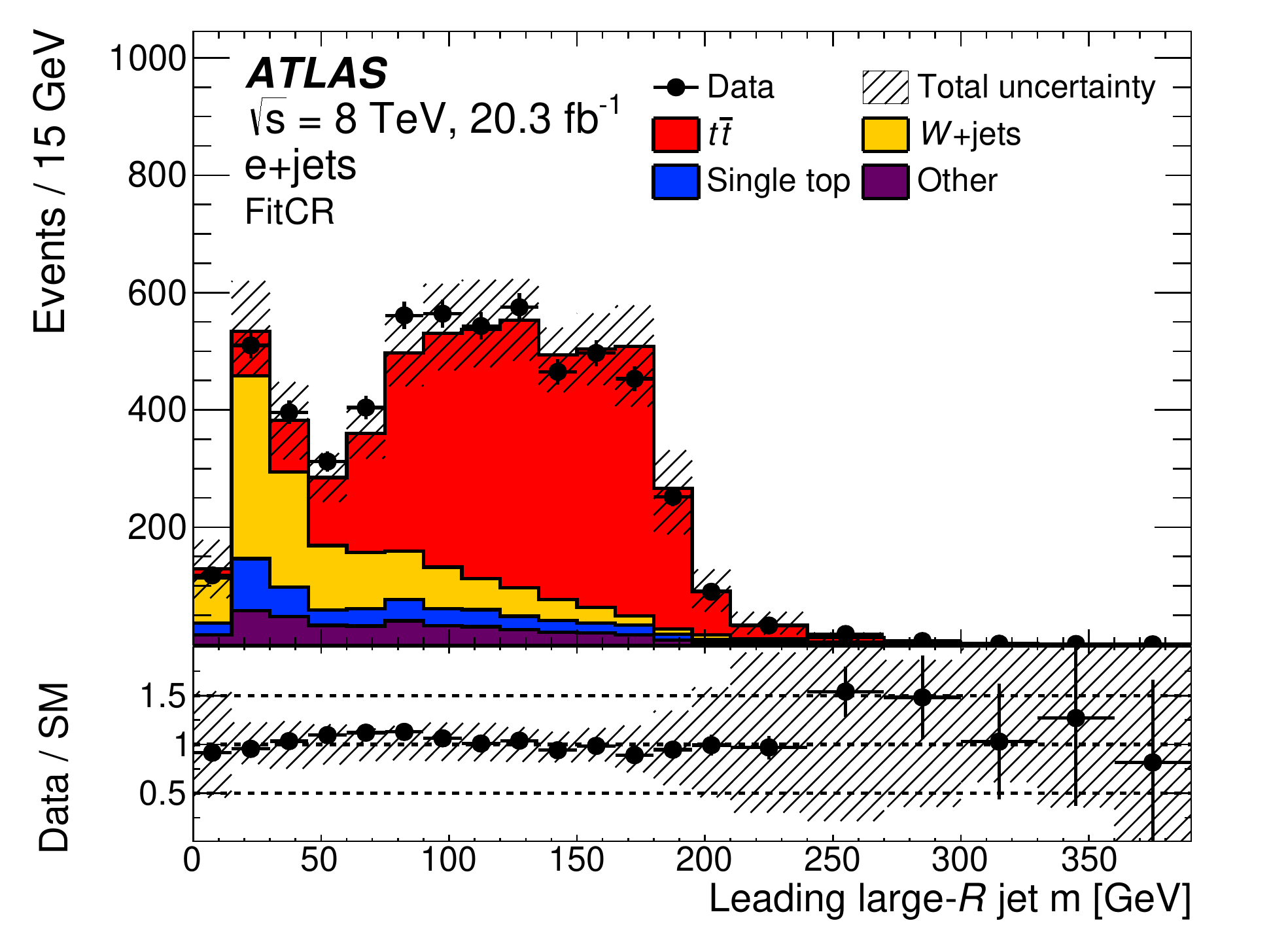}}
  \subfigure[$\mu$+jets channel]{\includegraphics[width=0.49\textwidth]{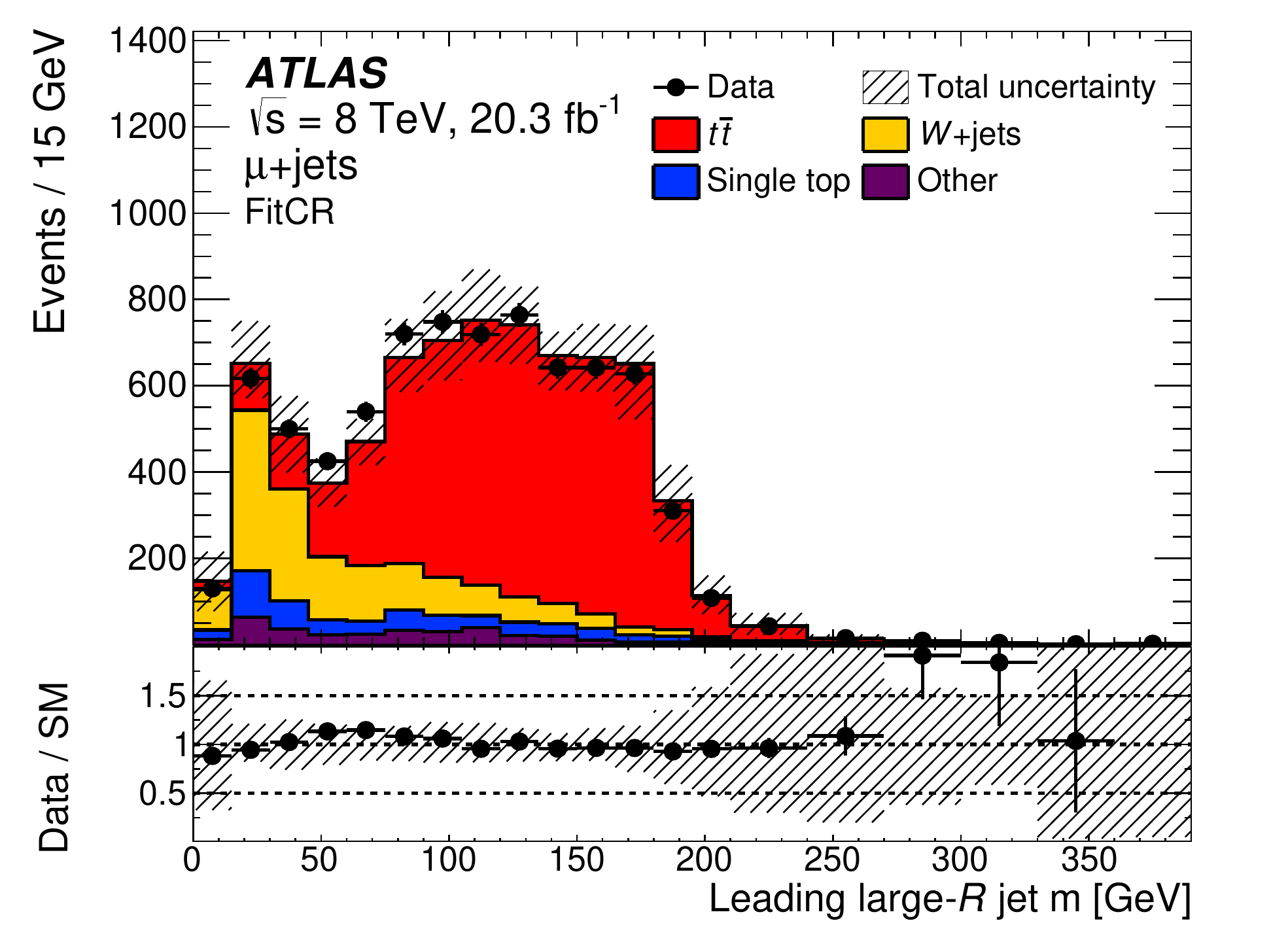}}
\caption{Comparison of data to the expected background for the leading large-$R$ jet mass in the FitCR, 
  both for the electron (left) and muon (right) channels, 
  after applying the $W$+jets and \ttbar\ normalisation correction factors.
}
\label{fig:fatjet_m_CR3}
\end{figure}

The \ttbar\ and $W$+jets normalisations are obtained from a fit to the large-$R$ jet mass distribution in the FitCR. 
The large-$R$ jet mass distribution for the $W$+jets contribution has a steeply falling shape, 
while the \ttbar\ fraction grows for values around the $W$-boson and top-quark masses. 
First, other small backgrounds, contributing less than 12\%, are subtracted from data. 
Normalisation correction factors are then obtained from the FitCR for the two background processes 
and the modelling is tested in the W1CR and the W2CR. 
Figure~\ref{fig:fatjet_m_CR3} shows the large-$R$ jet mass distribution in the FitCR,
including the corrections to the \ttbar\ and $W$+jets backgrounds.

The obtained correction factors with respect to the theoretical predictions for the muon (electron) channels are 0.874 (0.909) and
0.951 (0.947) for \Wboson+jets and \ttbar\ respectively.

After applying these corrections, a residual mismodelling 
of the $W$-boson \pT spectrum is observed at high \pT\ in all CRs.
To correct for this mismodelling, corrections are obtained in the FitCR and W2CR for both \ttbar\ and $W$+jets events
as a function of the $W$-boson \pT.
For \ttbar events, the derived correction factor is compatible with unity within the statistical uncertainties,
and is therefore not applied.
For $W$+jets, the correction factor is approximately unity for $W$-boson \pT below 300 \GeV,
decreasing to 0.6 for 500 \GeV\ and 0.4 for 600 \GeV.

\section{Analysis procedure}

\label{Sec:Analysis}

\begin{figure}
\subfigure[$e$+jets]{\includegraphics[width=0.49\textwidth]{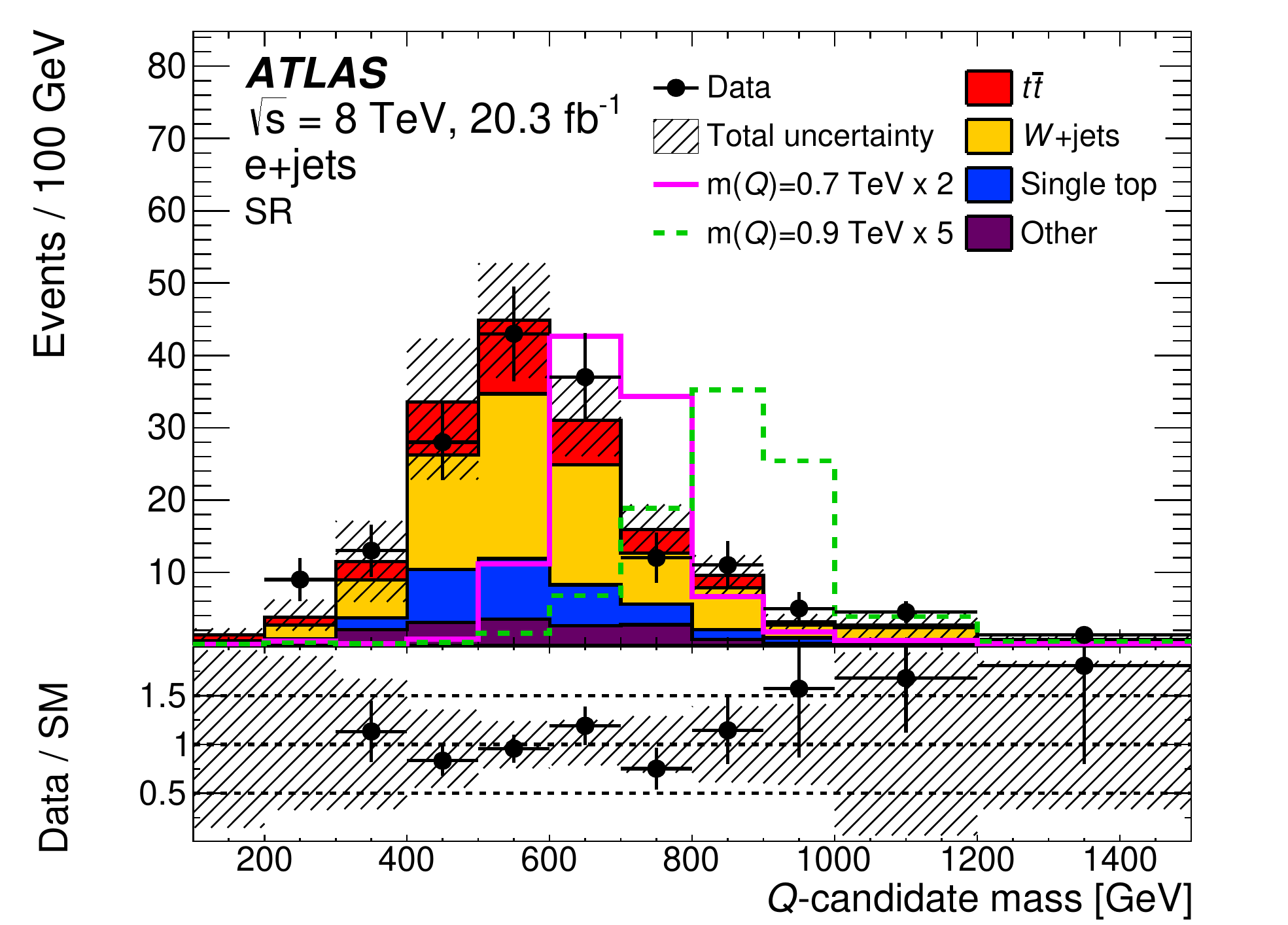}}
\subfigure[$\mu$+jets]{\includegraphics[width=0.49\textwidth]{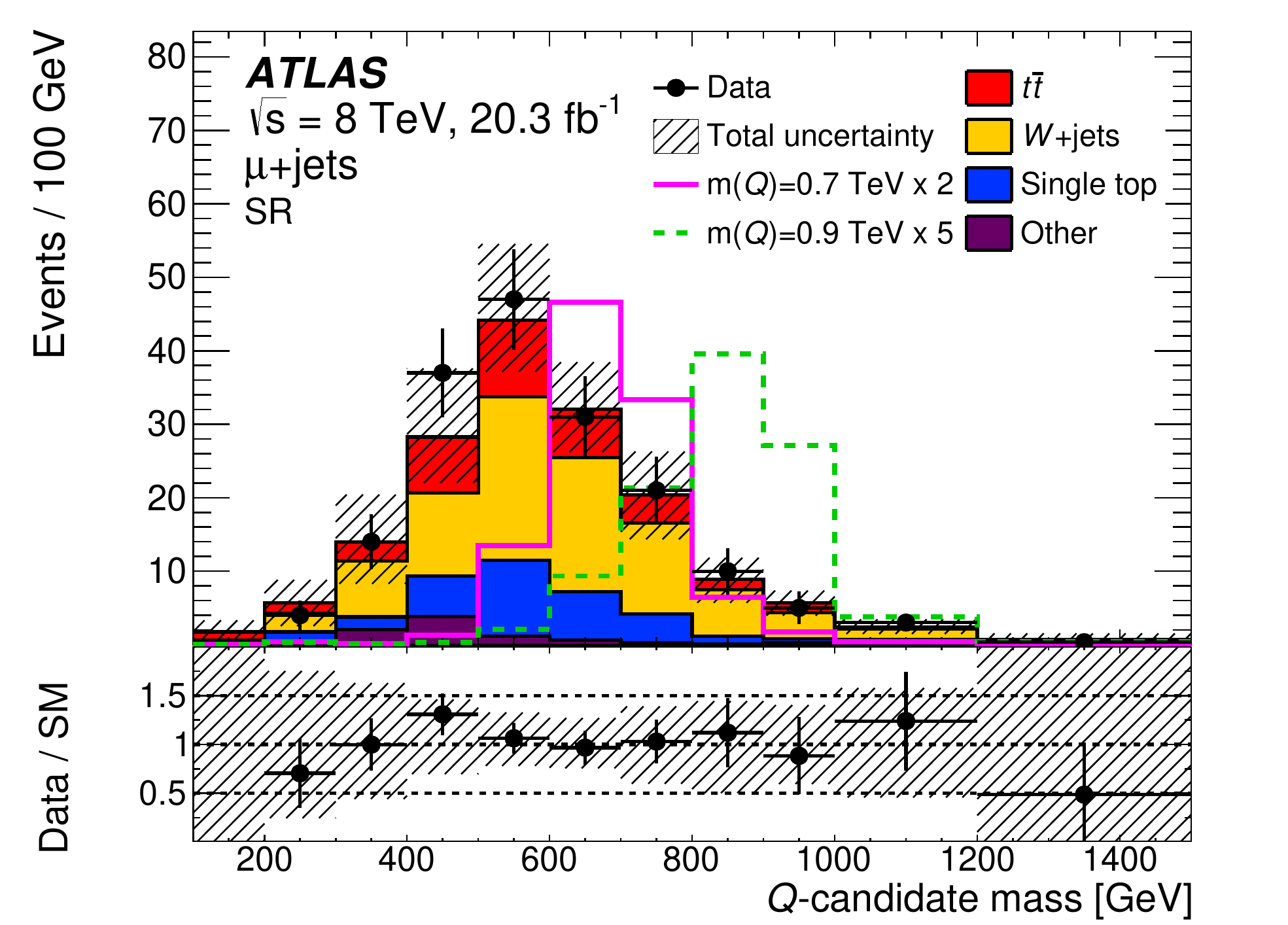}}
\caption{Distribution of the $Q$-candidate mass for the electron and muon channels before the likelihood fit. 
The signal yields are shown for cross-sections corresponding to \cL = 1 and for BR($Q \to  Wb$) = 0.5.
These are scaled up, in order to improve their visibility.
The uncertainty band includes all the uncertainties listed in Section~\ref{Systematics},
which are taken as fully uncorrelated between different sources.}
\label{fig:corr_c7_Tmass}
\end{figure}

After the event selection described in Section~\ref{Sec:EventSelection} and applying the correction factors obtained in 
Section~\ref{Sec:Background}, the $Q$ candidate is reconstructed. 
The first step is the reconstruction of the $W$-boson candidate by summing the four-momenta of the charged lepton and the neutrino.
To obtain the $z$-component of the neutrino momentum, the lepton-neutrino invariant mass is set to the $W$-boson mass
and the resulting quadratic equation is solved.
If no real solution exists, the \vecMET vector is varied by the minimum amount required to produce exactly one real solution.  
If two real solutions are found, the one with the smallest $|p_z|$ is used.
The $W$-boson candidate and the small-$R$ $b$-jet, which is matched to the large-$R$ jet, 
are then used to reconstruct the $Q$ candidate. 
Hence, no large-$R$ jet information is used directly for the reconstruction of the discriminant, 
which reduces the dependence of the final result on the systematic uncertainties of the large-$R$ jet kinematics. 
In Figure~\ref{fig:corr_c7_Tmass} the distribution of the $Q$-candidate mass in the SR is compared 
to the SM background prediction and the signal distributions for $m(Q) =$ 0.7 and \mbox{0.9 \TeV.} \\

A binned maximum-likelihood fit to the distribution of the $Q$-candidate mass is carried out using
the {\scshape HistFactory}~\cite{Cranmer:1456844} tool, which is part of the {\scshape HistFitter} \cite{Baak:2014wma} package. 
In the absence of signal, a profile-likelihood ratio is used to set an upper limit 
on the cross-section times BR at the 95\% CL. This is done using the CL$_s$ method~\cite{Junk:1999kv,0954-3899-28-10-313}. 
A combined fit to the electron and muon channels is performed. 
The systematic uncertainties are taken into account as nuisance parameters. 
The likelihood is then maximised using the nuisance parameters and the signal strength $\mu$ as parameters in the fit. 
The systematic uncertainty corresponding to each nuisance parameter is used as an a priori probability. 
These priors are assumed to follow a Gaussian distribution and constrain the nuisance parameters. 
The systematic uncertainties affecting both channels are treated as correlated across the channels.

\section{Systematic uncertainties}

\label{Systematics}

The shape and normalisation of the distribution of the $Q$-candidate mass is affected by various systematic uncertainties.
The sources of uncertainty are split into two categories: 
1) uncertainties due to the modelling of the signal 
and background processes; 2) experimental uncertainties on the calibration and efficiency for reconstructed objects. 
The impact of each source on the total signal and background normalisation is summarised in Table~\ref{Tab:Systematics}.

\begin{table}[!ht]
\centering
\caption{Summary of the impact of the systematic uncertainties on signal and background normalisations in percent. 
The values given for the signal are those corresponding to the 0.7 \TeV\ mass point.
If the uncertainties resulting from the up and down variations are asymmetric,
the larger deviation is shown here.}
\vspace{0.2cm}
\begin{tabular}{r|r|r}
\hline
\hline
 Systematic uncertainty & Signal & Total bkg. \\
\hline
\multicolumn{3}{c}{Modelling uncertainties [\%]} \\
\hline
$t\bar{t}$ and $W$+jets normalisation &  -  & 15    \\
$t\bar{t}$ modelling                  &  -  &  4.9  \\
$W$+jets modelling                    &  -  &  2.4  \\
Single top modelling                  &  -  &  6.3  \\
Multijet estimate                     &  -  &  2.6  \\
Parton distribution functions         & 2.0 &  7.4  \\
\hline
\multicolumn{3}{c}{Experimental uncertainties [\%]} \\
\hline
$b$-tagging                                & 8.0 & 1.5        \\
Small-$R$ jet energy resolution           & 0.7 & 0.3        \\
Small-$R$ jet energy scale                & 3.3 & 3.6        \\
JVF, small-$R$ jets                        & $< 0.1$ & 0.2    \\
Large-$R$ jet energy and mass resolution   & 4.0 & 6.8        \\
Large-$R$ jet energy scale                 & 7.2 & 9.7        \\
Lepton id \& reco                          & 2.3 & 0.2        \\
Missing transverse momentum                & 0.3 & 0.4        \\
Luminosity                                 & 2.8 & 2.7        \\
\hline
\hline
\end{tabular}
\label{Tab:Systematics}
\end{table}

\subsection{Modelling uncertainties}

The uncertainties are propagated from the FitCR to the SR, 
resulting in a background prediction uncertainty of 15\% in the SR
due to the statistical uncertainty in the FitCR. 
The \ttbar\ and $W$+jets normalisations are derived in the FitCR separately for 
each additional up and down variation accounting for a systematic uncertainty and applied in the SR. 
Therefore the uncertainties are taken to be fully correlated between the FitCR and SR. 

The uncertainties due to QCD initial- and final-state radiation modelling are estimated with samples generated 
with {\scshape AcerMC} interfaced to {\scshape Pythia6} for which the parton-shower parameters are varied according to a measurement 
of the additional jet activity in \ttbar\ events~\cite{ATLAS:2012al}.
The impact of the \ttbar\ modelling is evaluated using three different simulation samples described earlier in Section~\ref{Sec:Samples}. 
The uncertainty due to the choice of parton shower and hadronisation model
is evaluated by comparing samples produced with {\scshape Powheg+Pythia6} and {\scshape Powheg+Herwig}. 
For another comparison, the NLO matrix-element generator is changed simultaneously with the parton-shower model 
using samples generated with {\scshape Powheg}+{\scshape Pythia6} and {\scshape MC@NLO}+{\scshape Herwig}. 
Finally, the {\scshape Powheg+Pythia6} sample is compared to the LO sample generated with {\scshape Alpgen+Herwig}. 
The largest impact on the normalisation is observed when comparing {\scshape Powheg}+{\scshape Pythia6} and {\scshape MC@NLO}+{\scshape Herwig}. 
The total \ttbar\ modelling uncertainty is 4.9\%.

The dominant single-top-quark process is the $t$-channel production. 
In order to estimate the impact of using different models for this process, 
the nominal {\scshape Powheg+Pythia6} sample is compared to a sample produced with {\scshape MadGraph5\_}a{\scshape MC@NLO+Herwig}.
The change in the background acceptance is 6.3\%. 

To account for the shape uncertainties in the multijet background estimates, 
alternative methods are used in the evaluation of the real and fake rates for the matrix method. 
For the electron channel, the systematic uncertainties on the fake efficiencies are obtained by changing the parameterisation. 
For the muon channel, the fake efficiencies obtained in two different control regions are compared. 
The uncertainty on the real efficiency is estimated by comparing the values obtained from the tag-and-probe method 
with those from an alternative method, where very tight requirements are applied on \MET and $m_{\mathrm T}(W)$. 
An additional uncertainty is applied by varying the background normalisation in the control region for the fake estimate by 30\%, 
which corresponds to the uncertainty on the $Z$+jets and $W$+jets backgrounds in that control region. 
The resulting uncertainty on the background acceptance is 2.6\%.

To account for the mismodelling of the $W$-boson \pT, a polynomial fit is applied to obtain a continuous function for the reweighting.
This fit is repeated with different polynomials and the mean value of these functions is used as a nominal weight. 
Polynomials of degrees starting from one up to the maximum allowed by the number of degrees of freedom are used.
The largest deviation of the functions from the nominal weight in each bin is taken as a systematic uncertainty. 
The change in the background acceptance is 2.4\%.

To evaluate the PDF uncertainty, the uncertainties of three different PDF sets (NNPDF2.3 NLO~\cite{Ball:2012cx},
MSTW2008nlo~\cite{Martin:2009iq} and CT10 NLO) 
and their eigenvectors are considered. 
Based on the PDF4LHC recommendation~\cite{Botje:2011sn}, the envelope of all uncertainties is taken and symmetrised.
The resulting uncertainty on the background acceptance is 7.4\%.

\subsection{Experimental uncertainties} 

The detector response is affected by several sources of uncertainty which influence the object reconstruction 
and hence lead to a change in the selection efficiency. 
The effect on the signal yields is quoted for a $Q$ candidate with $m(Q) =$ 0.7 \TeV. 
In order to model the uncertainty on the $b$-jet identification, 
the $b$-tagging and mistagging scale factors are varied~\cite{Aad:2015ydr}. 
Large statistical fluctuations for high-momentum $b$-jets 
cause the corresponding systematic component to have a large impact on the total normalisation.
The $b$-tagging uncertainties affect the background by 1.5\% and the signal acceptance by 8\%. 
This difference arises because the impact of $b$-tagging uncertainties on the background is strongly mitigated 
by the use of the FitCR to determine the background normalisation.

The jet energy resolution is measured using in situ methods and the corresponding systematic uncertainty 
is about 10\% for jets with 30 $\leq p_\mathrm{T} \leq$ 500 \GeV~\cite{Aad:2012ag}. 
The total impact is 0.3\% on the background yields and 0.7\% on the signal yields. 
Pile-up suppression is achieved by applying a requirement on the JVF as described in Section~\ref{Sec:ObjectDefinition}.  
The JVF uncertainties affect the signal by < 0.1\% and the background yields by 0.2\%.

The small-$R$ jet energy scale~\cite{Aad:2014bia} uncertainty 
affects the background yield by 3.6\% and the signal acceptance by 3.3\%.
Although the large-$R$ jet is not directly used in the reconstruction of the $Q$ candidate, 
uncertainties related to the large-$R$ jet energy scale and resolution affect the event yields.
The uncertainty on the large-$R$ jet energy resolution and jet mass resolution results in an uncertainty of 
6.8\% on the background yield and an uncertainty of 4.0\% on the signal acceptance. 
The large-$R$ jet energy scale uncertainty has a larger effect: 9.7\% on the background acceptance and 7.2\% on the signal yield.
  
Uncertainties on trigger, reconstruction and identification efficiencies are evaluated in addition to uncertainties on lepton momentum scale and resolution. 
The impact of these uncertainties is $<$ 0.3\% on the background and 2.3\% on the signal acceptance. 
All experimental uncertainties are propagated consistently to the evaluation of the missing transverse momentum. 
The corresponding change in the event yields is smaller than 0.5\%. 
 
The uncertainty on the integrated luminosity is 2.8\%. 
It is derived, following the same methodology as that detailed in Ref.~\cite{Aad:2013ucp}.

\section{Results}
\label{sec:result}

The event yields obtained in the signal region for an integrated luminosity of 20.3 fb$^{-1}$ are displayed in Table~\ref{table:yield_c7}. 
The expected yields for signal masses of 0.7 and 0.9 \TeV\ are shown alongside the background prediction,
which includes the normalisation of the \ttbar and $W$+jets event yields obtained in the FitCR and the number of events observed in data. 

\begin{table}[ht]
\centering
\caption{
      Comparison of the observed number of events with the expected number before the fit
      in the signal region after applying the corrections and the full event selection. 
      The normalisation of the \ttbar and $W$+jets backgrounds was obtained in the FitCR.  
      The statistical and systematic uncertainties on the MC predictions are presented here and are symmetrised. 
      The signal yields are shown for \cL$=1$ and BR($T \rightarrow Wb$) = 0.5. 
}
\vspace{0.2cm}
\begin{tabular}[c]{cll}
\hline
\hline  &  $e$+jets & $\mu$+jets \\ 
\hline
$T$ (0.7 \TeV) & $\phantom{0}$50$\phantom{.1}$ $\pm$ $\phantom{0}$7$\phantom{.1}$   & $\phantom{0}$52$\phantom{.1}$ $\pm$ $\phantom{0}$7 \\
$T$ (0.9 \TeV) & $\phantom{0}$19.6 $\pm$ $\phantom{0}$3.3   & $\phantom{0}$21.8 $\pm$ $\phantom{0}$3.4 \\
\hline

$W$+jets       & $\phantom{0}$82$\phantom{.1}$ $\pm$ 28$\phantom{.1}$               & $\phantom{0}$89$\phantom{.1}$ $\pm$ 33 \\
\ttbar\        & $\phantom{0}$34$\phantom{.1}$ $\pm$ 27$\phantom{.1}$               & $\phantom{0}$37$\phantom{.1}$ $\pm$ 30 \\
Single top     & $\phantom{0}$29$\phantom{.1}$ $\pm$ 19$\phantom{.1}$               & $\phantom{0}$33$\phantom{.1}$ $\pm$ 15 \\
\Zboson+jets   & $\phantom{00}$6$\phantom{.1}$ $\pm$  $\phantom{0}$4$\phantom{0.1}$ & $\phantom{00}$4$\phantom{.1}$ $\pm$ $\phantom{0}$4 \\
Diboson        & $\phantom{00}$3$\phantom{.1}$ $\pm$  $\phantom{0}$1$\phantom{0.1}$ & $\phantom{00}$2$\phantom{.1}$ $\pm$ $\phantom{0}$1 \\
Multijets      & $\phantom{00}$8$\phantom{.1}$ $\phantom{0}^{\phantom{0}+12}_{\phantom{0}-8}$ & $\phantom{00}$3.2 $\pm$ $\phantom{0}$1.2\\[1.05ex]

\hline
SM bkg.        & 162$\phantom{.1}$ $\pm$ 43$\phantom{.1}$                           & 168$\phantom{.1}$ $\pm$ 46 \\
\hline Data    & 171$\phantom{0}$                                                   & 176$\phantom{0}$ \\ 
\hline
\hline
\end{tabular}
\label{table:yield_c7}\end{table}

\begin{figure}[!htb]
\subfigure[$e$+jets]{\includegraphics[width=0.49\textwidth]{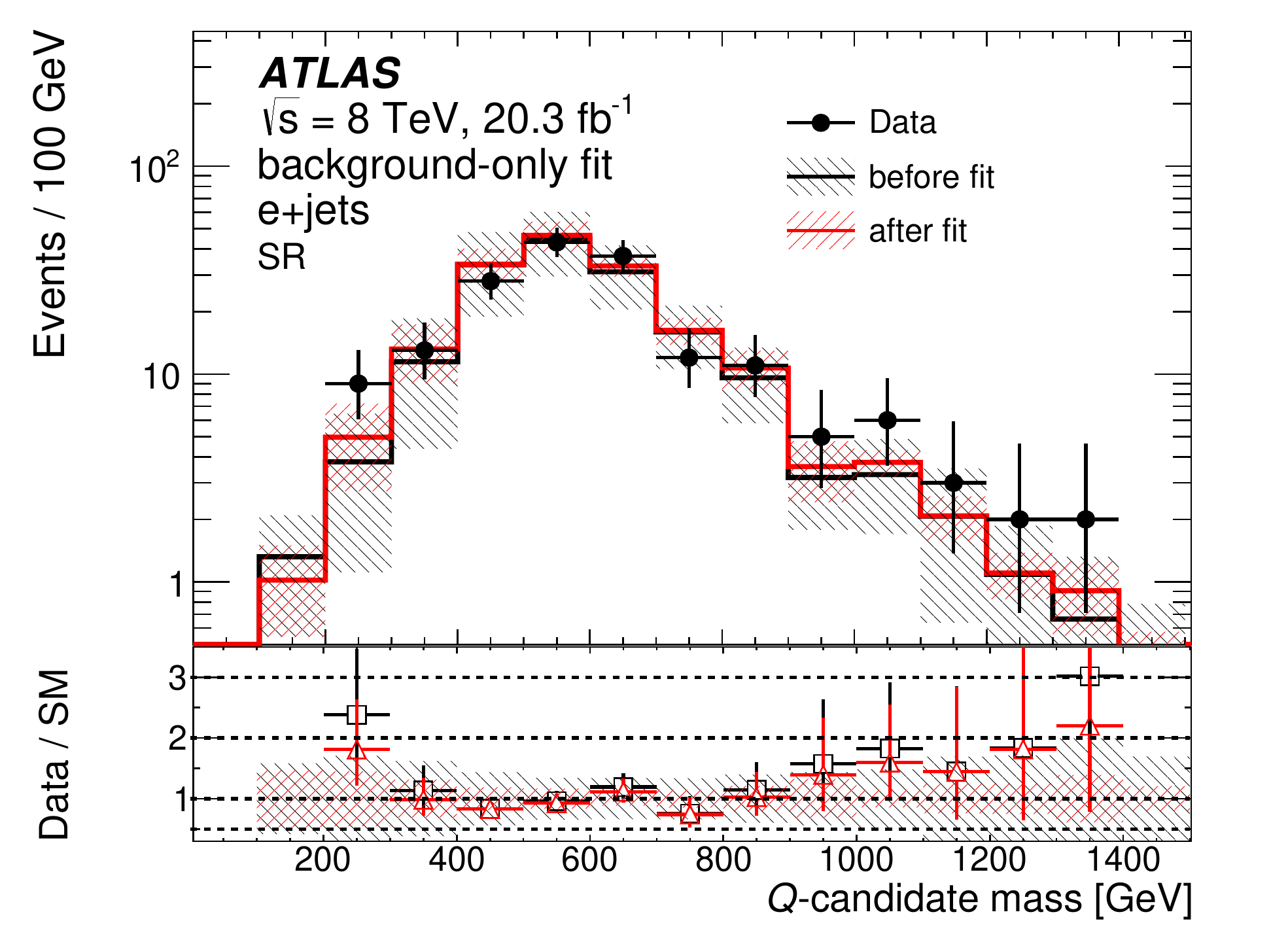}}
\subfigure[$\mu$+jets]{\includegraphics[width=0.49\textwidth]{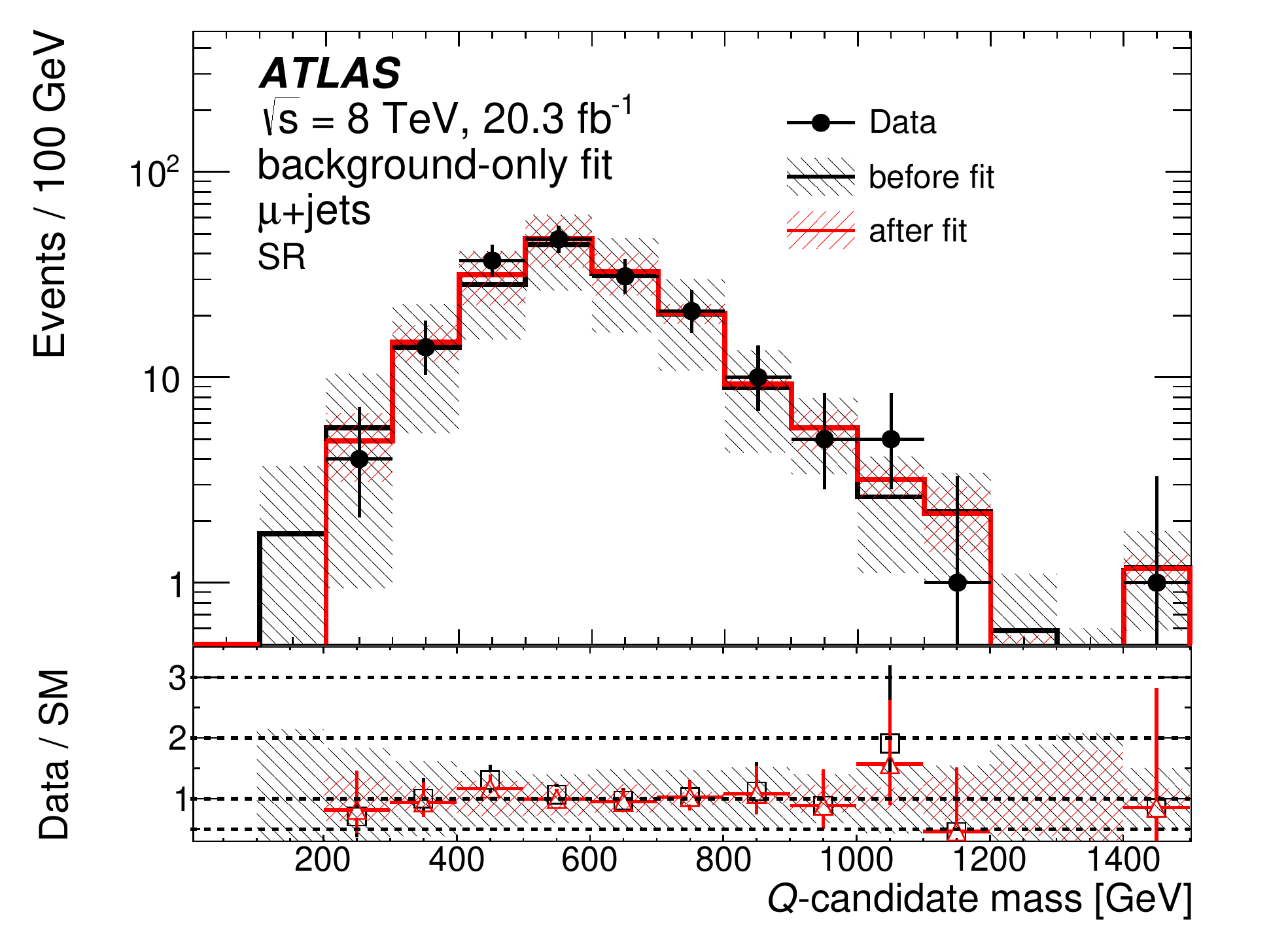}}
\caption{Distribution of the $Q$-candidate mass for the electron (left) and muon (right) channels, 
both before and after the nuisance parameter fit. The fit was performed using a background-only hypothesis. 
The error bands include the full statistical and systematic uncertainty before and after the fit.
The bottom panels show the ratio between the observed data and the SM prediction before (black squares)
and after (red triangles) the nuisance parameter fit.}
\label{fig:SMbkgr_beforeVSpostFit}
\end{figure}

No significant deviation from the SM background prediction is found. 
In the electron channel there is a tendency for the number of events in data 
to exceed the expectation for candidate masses above 0.9 \TeV.  
The  local $p_0$-value for the observed data to agree with the background-only hypothesis 
reaches its smallest value of 5.2\% (corresponding to 1.6 standard deviations) at 1 \TeV.
Mass-dependent exclusion limits in steps of 0.1 \TeV\ are set 
on the cross-section times BR of the $Q$ candidate as explained in Section~\ref{Sec:Analysis}. 
A simultaneous maximum-likelihood fit is performed to the electron and muon distributions. 
In Figure~\ref{fig:SMbkgr_beforeVSpostFit} the mass distributions before (black) and after (red) 
the nuisance parameter fit (background-only hypothesis) are compared. 
The narrower uncertainty band for the post-fit distribution shows that the overall uncertainty is reduced in the nuisance parameter fit. 

\begin{figure}[!htb]
\centering
\includegraphics[width=0.5\textwidth]{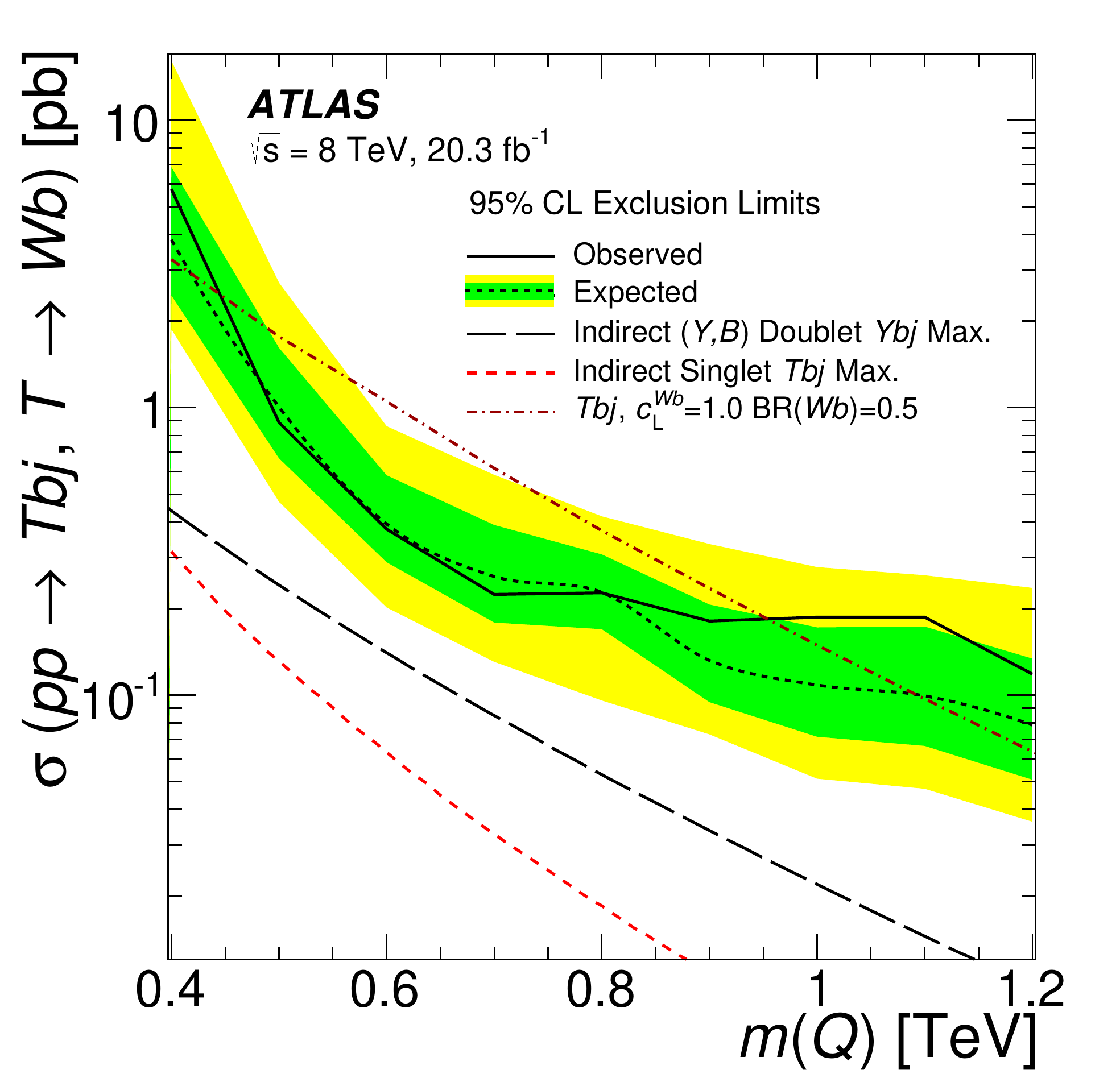}
\caption{Observed and expected limits on the cross-section times BR 
for the single production of a vector-like quark $Q \to Wb$ as a function of the $Q$ mass.
The limits are shown compared to three theoretical predictions: the NLO cross-section prediction
in the composite-Higgs-model scenario~\cite{Matsedonskyi:2014mna} (brown dot-dashed line), 
and the maximum cross-sections for $Tbj$ (red dashed line) and $Ybj$ (black dashed line)~\cite{Aguilar-Saavedra:2013qpa}.
}
\label{fig:Limit_XsecBR}
\end{figure}

The observed and expected 95\% CL limits on the cross-section times BR of singly produced 
$Q$ candidates is shown in Figure~\ref{fig:Limit_XsecBR} for different candidate masses.
The expected upper limit on the cross-section is determined using pseudo-data constructed 
from a background-only model built from the nuisance parameters fitted to real data.
The limits include full statistical and systematic uncertainties and are compared to the maximum allowed cross-sections for $Tbj$ and $Ybj$ 
from electroweak constraints~\cite{Aguilar-Saavedra:2013qpa} 
and the NLO cross-section prediction for $\cL=1$~\cite{Matsedonskyi:2014mna}.
The observed direct limits are less stringent than the indirect limits 
on the maximum cross-sections from Ref.~\cite{Aguilar-Saavedra:2013qpa}, 
but rely on fewer assumptions about the new physics that would produce $T$ or $Y$ quarks.

More events than predicted are observed for the higher mass values, leading to a less stringent observed limit for masses above 0.8 \TeV. 
These differences are, however, within the 1$\sigma$ uncertainty band. 
The mass limit is obtained from the intersection of the NLO prediction with the curve for the observed cross-section times BR limit. 
The observed (expected) limit on the $Q$-candidate mass obtained for this scenario is 0.95 (1.10) \TeV.

\subsection{Interpretation for singlet vector-like $T$ quarks}

\begin{figure}
\centering
\subfigure[]{\includegraphics[width=0.49\textwidth]{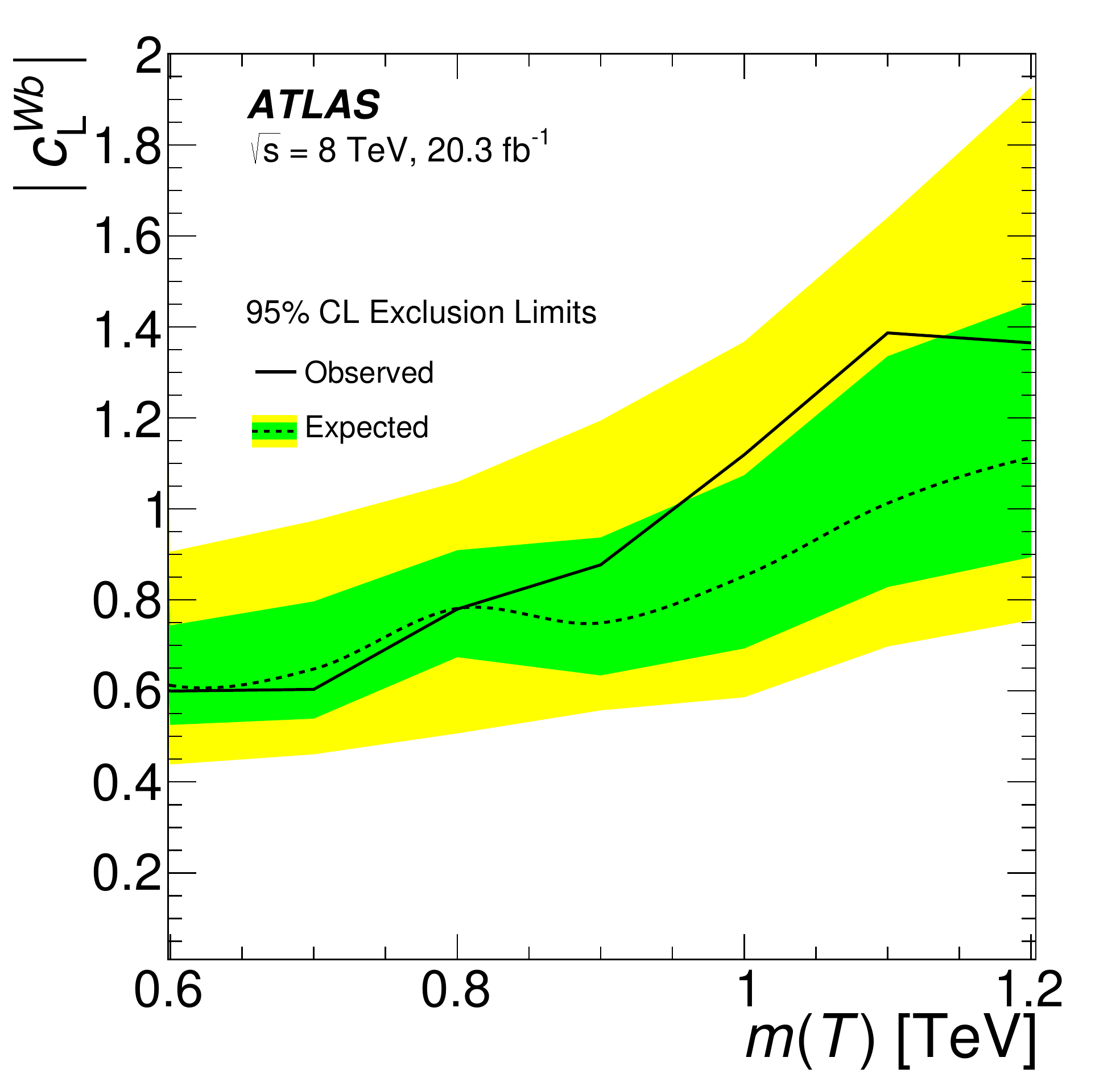}}
\subfigure[]{\includegraphics[width=0.49\textwidth]{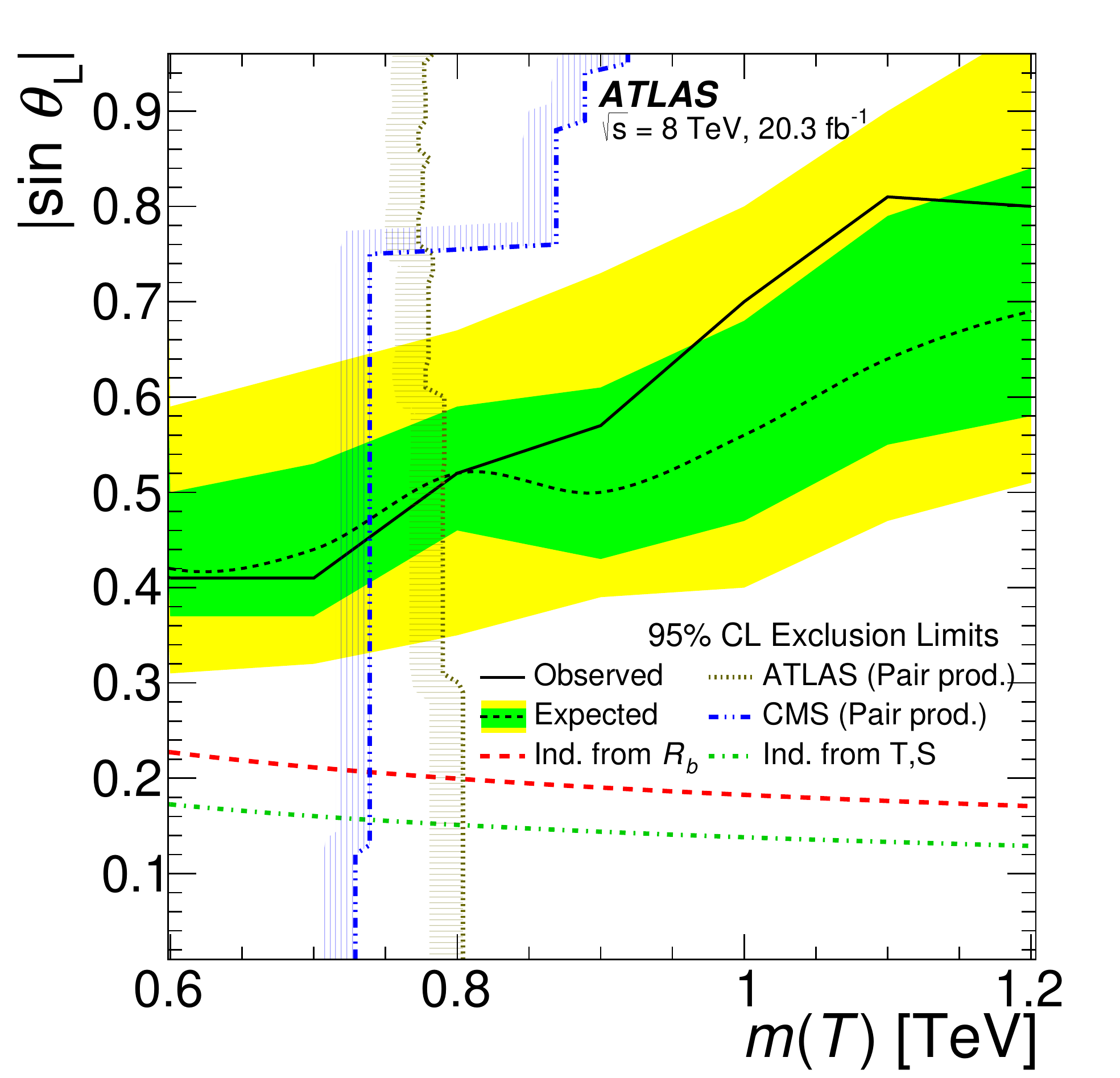}}
\caption{(a) Observed and expected limit (95\% CL) on the coupling of the vector-like quark to the SM \Wboson\ boson and $b$-quark
as a function of the $Q$ mass,
where the BR($T\rightarrow Wb$) is assumed to be 50\%.
The excluded region is given by the area above the solid black line. 
(b) Observed and expected limit (95\% CL) on the mixing of a singlet vector-like $T$ quark to the SM sector,
where the BR($T\rightarrow Wb$) is assumed to be that of a singlet.
The excluded region is given by the area above the solid black line. 
The limits are shown compared to the indirect electroweak constraints from Ref.~\cite{Aguilar-Saavedra:2013qpa} (green and red line).
In addition, the observed limits from pair-production searches by ATLAS~\cite{Aad:2015kqa} (olive) 
and CMS~\cite{Khachatryan:2015oba} (blue) are shown.
}
\label{fig:IngredientsJA}
\end{figure}

The limit set on the cross-section times branching ratio can be translated into a limit on \cL, using the relation
\begin{align}
|\cL| &= \sqrt{\frac{\sigma_{\mathrm limit}}{\sigma_{\mathrm theory}}}
\end{align}
and the theoretical predictions from Ref.~\cite{Matsedonskyi:2014mna}. For the theoretical prediction the value of \cL\ was set to 1.0. 
The expected and observed limits are shown in Figure~\ref{fig:IngredientsJA}(a).
These limits exclude couplings above 0.6 for masses below 0.7 \TeV\ and above \cL= 1.2 for a $T$ quark with a mass of 1.2~\TeV. 
The limits on the mixing angle between the vector-like quark and the SM sector are derived in a similar fashion
and are shown in Figure~\ref{fig:IngredientsJA}(b). 
For lower masses, mixing angles from 0.4 to 0.5 are excluded, while the limit increases up to 0.81 for a $T$ quark with a mass of 1.2~\TeV. 

As shown in Formula B1 of Ref.~\cite{Matsedonskyi:2014mna}, 
the width of the vector-like quark is proportional to \cTwoL. 
Therefore, a larger width is expected for higher values of \cL. As described in Section~\ref{Sec:Samples}, 
a narrow-width approximation is used in the production of the signal samples. 
To test the validity of the limits shown in Figure~\ref{fig:IngredientsJA}, 
the limits were recalculated for signal samples with $\Gamma/m$ values up to 0.46,
using the same theoretical cross-section prediction.
For all masses and $\Gamma/m$ the observed limit is found to be more stringent than, or equal to, 
the value obtained for the narrow-width approximation.
For $m(Q)=0.9$ \TeV\ the cross-section times BR limit decreases by 15\% (20\%) for $\Gamma/m =$ 0.3 ($\Gamma/m =$ 0.46)
and for $m(Q)=1.2$ \TeV\ the limit decreases by 13\% (21\%) for $\Gamma/m =$ 0.3 ($\Gamma/m= $ 0.46).
Hence, the limits presented in this paper constitute a conservative estimate
regarding the assumptions about the width of vector-like quarks.

\subsection{Interpretation for a vector-like $Y$ quark from a doublet}

\begin{figure}[!htb]
\centering
\subfigure[]{\includegraphics[width=0.49\textwidth]{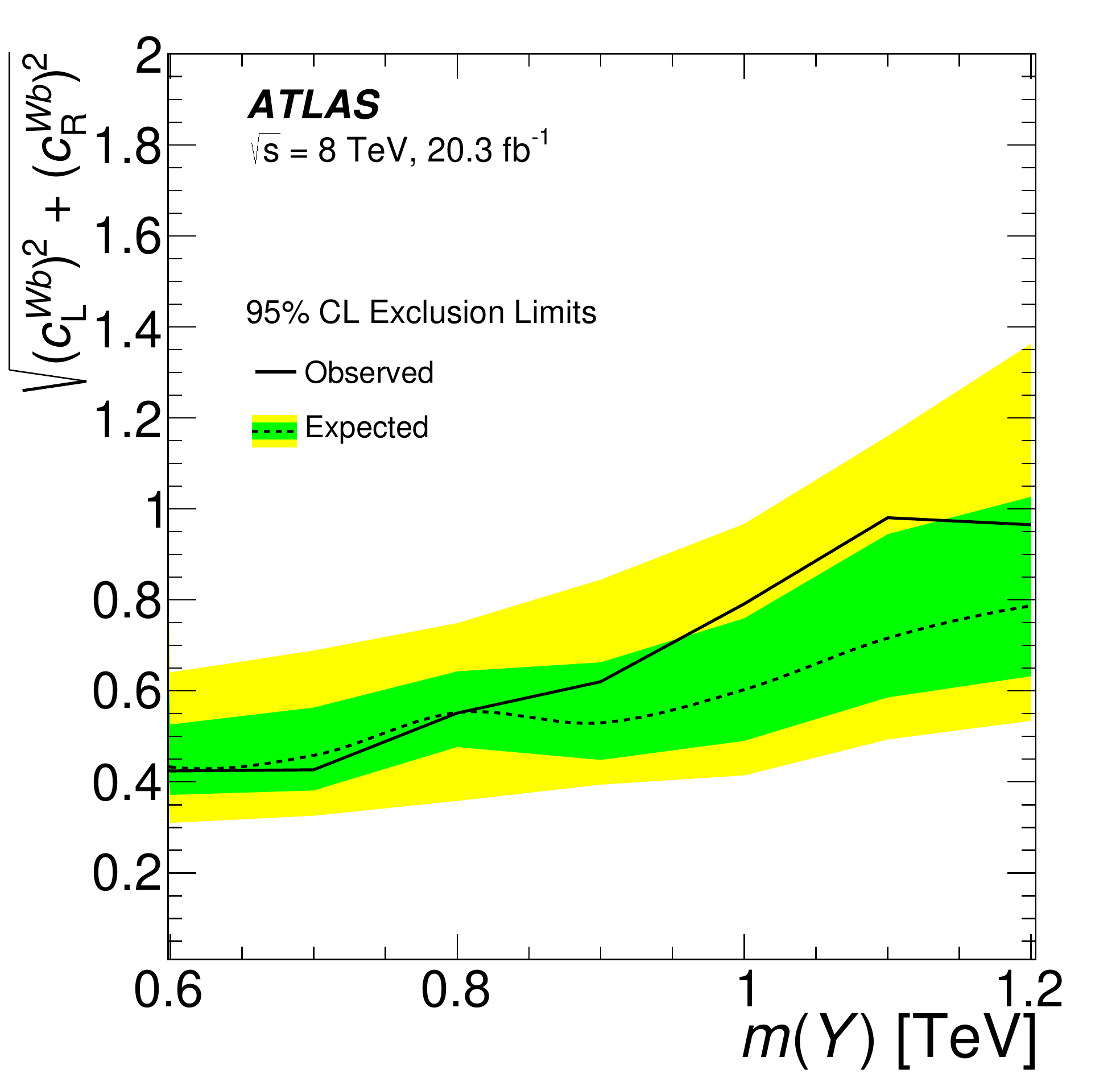}}
\subfigure[]{\includegraphics[width=0.49\textwidth]{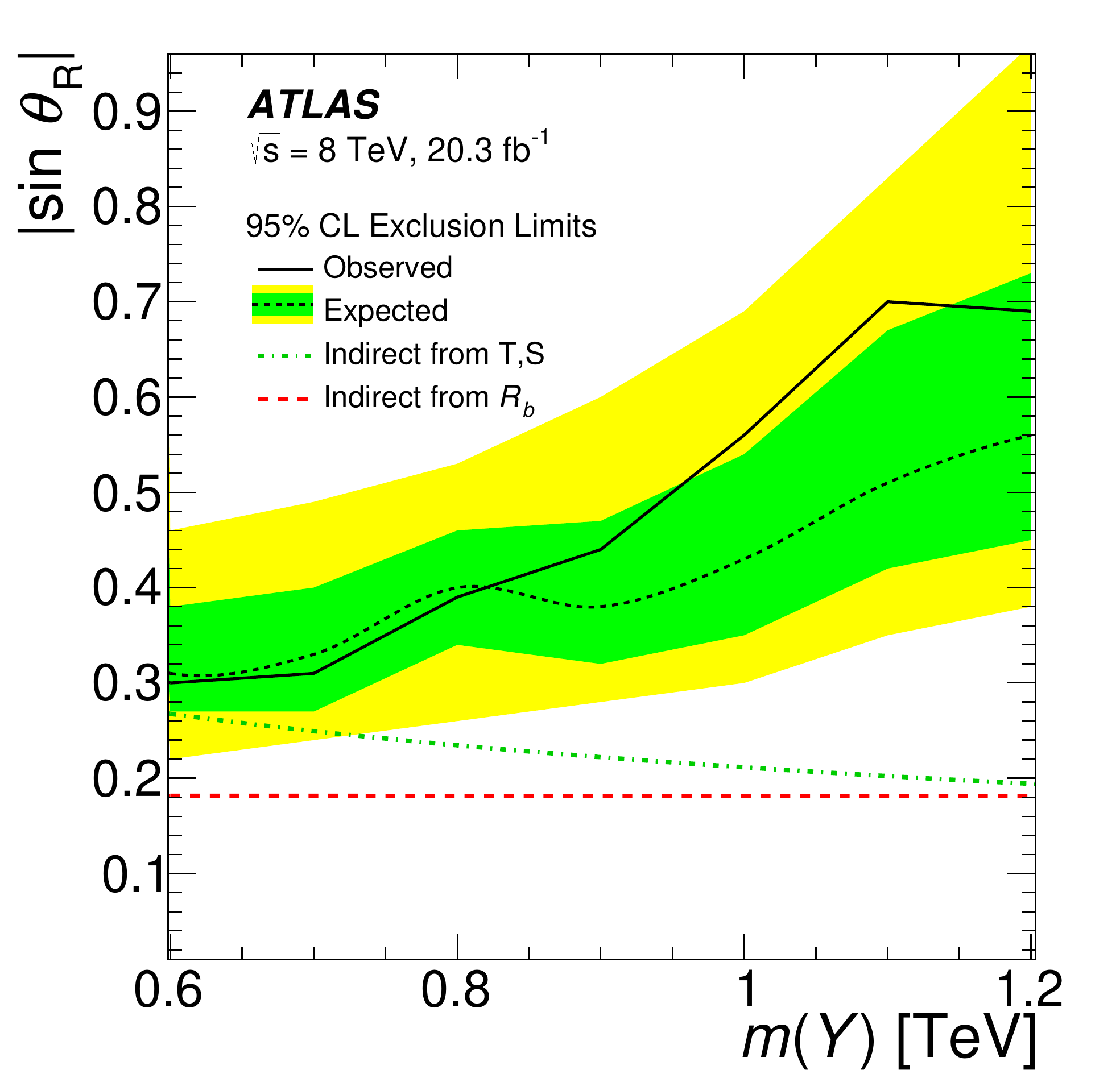}}
\caption{(a) Observed and expected 95\%\ CL upper limits on the coupling of the vector-like $Y$ quark 
to the SM \mbox{\Wboson\ boson} and $b$-quark as a function of the $Q$ mass.
(b) Observed and expected 95\%\ CL upper limits on the mixing of a vector-like $Y$ quark to the SM sector in a $(Y,B)$ doublet model. 
In addition, the indirect electroweak constraints from Ref.~\cite{Aguilar-Saavedra:2013qpa} are shown.
For both (a) and (b) BR($Y\rightarrow Wb$) is assumed to be 100\%.}
\label{fig:JA_sL}
\end{figure}

The limits on cross-section times BR are used to set limits on the couplings \cL and \cR for a vector-like $Y$ quark. 
Using the theoretical cross-section and the general vector-like quark model discussed in Ref.~\cite{Matsedonskyi:2014mna} 
as well as the BR($Y \to Wb$) = 1, a limit on $\sqrt{\cTwoL+\cTwoR}$ is set. 
Due to the higher BR of the vector-like $Y$ quark, this limit as shown in Figure~\ref{fig:JA_sL}(a) 
is more stringent, by a factor of $1/\sqrt{2}$, than the limit on $|\cL|$ for single $T$ production.
The cross-section limit is also translated into a limit on the mixing parameter $|\sinR|$ in a $(Y,B)$ doublet model. 
This is done as a function of the $Y$ mass as discussed in Ref.~\cite{Aguilar-Saavedra:2013qpa}. 
Figure~\ref{fig:JA_sL}(b) shows the resulting limit on $|\sinR|$ for the $(Y,B)$ doublet as a function of $m(Y)$, 
including also the limit on $|\sinR|$ for a $(Y,B)$ doublet model from electroweak precision observables 
taken from Ref.~\cite{Aguilar-Saavedra:2013qpa}.

\clearpage
\section{Summary}
A search for the production of a single vector-like quark $Q$ with subsequent decay into $Wb$ 
has been carried out with the ATLAS experiment at the LHC. 
The data used in this search correspond to 20.3 \ifb of $pp$ collisions at a centre-of-mass energy of $\sqrt{s}$ = 8 \TeV. 
The selected events have exactly one isolated electron or muon, at least two small-$R$ jets, at least one large-$R$ jet, 
one $b$-tagged jet and missing transverse momentum. 
Events with massive large-$R$ jets are vetoed to reduce the $t\bar{t}$ and $W$+jets background processes.
The $Q$ candidate is fully reconstructed and its mass is used as discriminating variable in a maximum-likelihood fit. 
The observed data distributions are compatible with the Standard Model background prediction and no significant excess is observed. 
Upper limits are set on the cross-section times branching ratio as a function of the $T$-quark mass using \cL= 1 and BR($T \to Wb$) = 0.5. 
The observed (expected) exclusion limit for $T$ quarks is 0.95 \TeV\ (1.10 \TeV) at the 95\% confidence level.
Using theoretical predictions, the cross-section limits are translated 
into limits on the $QWb$ coupling \cL and the mixing angle of the $T$ quark with the SM sector. 
The results are also interpreted as the coupling of a vector-like $Y$ quark to the SM $W$~boson and $b$-quark 
as well as a limit on the mixing parameter $|\sinR|$ in a $(Y,B)$ doublet model.

\label{sec:summary}
\clearpage
\section*{Acknowledgements}

We thank CERN for the very successful operation of the LHC, as well as the
support staff from our institutions without whom ATLAS could not be
operated efficiently.

We acknowledge the support of ANPCyT, Argentina; YerPhI, Armenia; ARC, Australia; BMWFW and FWF, Austria; ANAS, Azerbaijan; SSTC, Belarus; CNPq and FAPESP, Brazil; NSERC, NRC and CFI, Canada; CERN; CONICYT, Chile; CAS, MOST and NSFC, China; COLCIENCIAS, Colombia; MSMT CR, MPO CR and VSC CR, Czech Republic; DNRF and DNSRC, Denmark; IN2P3-CNRS, CEA-DSM/IRFU, France; GNSF, Georgia; BMBF, HGF, and MPG, Germany; GSRT, Greece; RGC, Hong Kong SAR, China; ISF, I-CORE and Benoziyo Center, Israel; INFN, Italy; MEXT and JSPS, Japan; CNRST, Morocco; FOM and NWO, Netherlands; RCN, Norway; MNiSW and NCN, Poland; FCT, Portugal; MNE/IFA, Romania; MES of Russia and NRC KI, Russian Federation; JINR; MESTD, Serbia; MSSR, Slovakia; ARRS and MIZ\v{S}, Slovenia; DST/NRF, South Africa; MINECO, Spain; SRC and Wallenberg Foundation, Sweden; SERI, SNSF and Cantons of Bern and Geneva, Switzerland; MOST, Taiwan; TAEK, Turkey; STFC, United Kingdom; DOE and NSF, United States of America. In addition, individual groups and members have received support from BCKDF, the Canada Council, CANARIE, CRC, Compute Canada, FQRNT, and the Ontario Innovation Trust, Canada; EPLANET, ERC, FP7, Horizon 2020 and Marie Sk{\l}odowska-Curie Actions, European Union; Investissements d'Avenir Labex and Idex, ANR, R{\'e}gion Auvergne and Fondation Partager le Savoir, France; DFG and AvH Foundation, Germany; Herakleitos, Thales and Aristeia programmes co-financed by EU-ESF and the Greek NSRF; BSF, GIF and Minerva, Israel; BRF, Norway; the Royal Society and Leverhulme Trust, United Kingdom.

The crucial computing support from all WLCG partners is acknowledged
gratefully, in particular from CERN and the ATLAS Tier-1 facilities at
TRIUMF (Canada), NDGF (Denmark, Norway, Sweden), CC-IN2P3 (France),
KIT/GridKA (Germany), INFN-CNAF (Italy), NL-T1 (Netherlands), PIC (Spain),
ASGC (Taiwan), RAL (UK) and BNL (USA) and in the Tier-2 facilities
worldwide.

\clearpage
\printbibliography
\clearpage
\begin{flushleft}

\bigskip

G.~Aad$^\textrm{\scriptsize 85}$,
B.~Abbott$^\textrm{\scriptsize 112}$,
J.~Abdallah$^\textrm{\scriptsize 150}$,
O.~Abdinov$^\textrm{\scriptsize 11}$,
B.~Abeloos$^\textrm{\scriptsize 116}$,
R.~Aben$^\textrm{\scriptsize 106}$,
M.~Abolins$^\textrm{\scriptsize 90}$,
O.S.~AbouZeid$^\textrm{\scriptsize 157}$,
H.~Abramowicz$^\textrm{\scriptsize 152}$,
H.~Abreu$^\textrm{\scriptsize 151}$
% ...
\bigskip
\\
% ...
$^{11}$ Institute of Physics, Azerbaijan Academy of Sciences, Baku, Azerbaijan\\
% ...
$^{85}$ CPPM, Aix-Marseille Universit{\'e} and CNRS/IN2P3, Marseille, France\\
% ...
$^{90}$ Department of Physics and Astronomy, Michigan State University, East Lansing MI, United States of America\\
% ...
$^{106}$ Nikhef National Institute for Subatomic Physics and University of Amsterdam, Amsterdam, Netherlands\\
% ...
$^{112}$ Homer L. Dodge Department of Physics and Astronomy, University of Oklahoma, Norman OK, United States of America\\
% ...
$^{116}$ LAL, Univ. Paris-Sud, CNRS/IN2P3, Universit{\'e} Paris-Saclay, Orsay, France\\
% ...
$^{150}$ Institute of Physics, Academia Sinica, Taipei, Taiwan\\
$^{151}$ Department of Physics, Technion: Israel Institute of Technology, Haifa, Israel\\
$^{152}$ Raymond and Beverly Sackler School of Physics and Astronomy, Tel Aviv University, Tel Aviv, Israel\\
% ...
$^{157}$ Department of Physics, University of Toronto, Toronto ON, Canada\\
% ...
\end{flushleft}


\end{document}